\documentclass[
print,            
amsmath,amssymb,    
aps,                
prd,                
]{revtex4-2}

\usepackage{graphicx}
\usepackage{dcolumn}
\usepackage{bm}
\usepackage{multirow}
\usepackage[colorlinks=true,citecolor=blue,urlcolor=blue]{hyperref}
\begin{document}
	
	\preprint{APS/123-QED} 
	
	\title{$I-$Love$-$Curvature: Exploring compact stars' quasi-universal relation with curvature scalars}
	\author{M. D. Danarianto$^{a}$}
	\email{m.dio.danarianto@brin.go.id}
	\author{A. Sulaksono$^b$}
	\email{anto.sulaksono@sci.ui.ac.id}
	
	\affiliation{$^a$Research Center for Computing, National Research and Innovation Agency (BRIN), Bandung 40173, Indonesia}
	
	\affiliation{$^b$Departemen Fisika, FMIPA Universitas Indonesia, Kampus UI, Depok 16424, Indonesia}
	
	\date{\today}
	
	\begin{abstract}
		We investigate quasi-universal relations in neutron stars linking standard observables, such as tidal deformability ($\Lambda$) and normalized moment of inertia ($\bar{I}$), with normalized curvature scalars in general relativity. These curvature scalars include the Ricci scalar ($\mathcal{R}$), the Ricci tensor contraction ($\mathcal{J}$), the Weyl scalar ($\mathcal{W}$), and the Kretschmann scalar ($\mathcal{K}$). We systematically examine both piecewise polytropic and color-flavor-locked equations of state, finding: (1) significant correlations between both local (central and surface) and global (volume-averaged) curvature scalars with $\bar{I}$ and $\Lambda$; (2) especially strong correlations between surface and volume-averaged curvature scalars and both $\bar{I}$ and $\Lambda$; (3) a near equation-of-state-independent maximum for the normalized Ricci scalar, suggesting a link to the trace anomaly; and (4) new universal relations involving normalized central and volume-averaged pressure and energy density, which also correlate strongly with $\bar{I}$ and $\Lambda$. Using constraints from GW170817 and low-mass X-ray binaries, we demonstrate that $\Lambda$ measurements directly constrain both scalar curvature quantities and the interior properties of canonical-mass neutron stars. These findings agree with the literature on equation-of-state-dependent Bayesian inference estimates. Our identified relations thus provide an equation-of-state-insensitive connection between stellar observables, spacetime geometry, and the microphysics of compact stars.
	\end{abstract}
	
	\maketitle
	
	
	\section{Introduction}
	The equation of state (EoS) for compact objects, including neutron stars and quark stars, remains highly uncertain. Universal relations among the properties of compact objects are essential. They provide model-independent methods for analyzing and predicting these properties. Universal relations also constrain properties of compact objects that are challenging or impossible to observe directly. They facilitate the integration of diverse observational data for a more comprehensive understanding. For a detailed review of universal relations, see Ref.~\cite{Yagi2017}.
	
	The quasi-universal relation between dimensionless tidal deformability and moment of inertia, as described in Ref.~\cite{Yagi2013a}, is both theoretically significant and astrophysically relevant. It links multiple neutron star observables. These equation-of-state insensitive relations—particularly the $I$--Love--$Q$ relation \cite{Yagi2013} and its extensions—facilitate the connection of properties such as asteroseismological oscillation modes \cite{Lau2009,Sotani2021}, interior structure indicators including $P_c/\varepsilon_c$ and $\langle c_s^2 \rangle$ \cite{Saes2021,Saes2024}, and spacetime structure descriptors such as surface redshift and compactness \cite{Yang2022}. Surface Kretschmann scalar \cite{Das2023} and the membrane paradigm \cite{Silvestrini2025} are also recently studied. In addition, strong correlations between quantities such as $I$ and $\Lambda$ provide further links to features like the rotation-induced quadrupole moment and $f$-mode oscillations. These relations enable the integration of observational constraints across diverse stellar properties, as demonstrated in Ref.~\cite{Kumar2019}.

	From a theoretical standpoint, universal relations highlight basic features of general relativity (GR) \cite{Yagi2013}. The underlying theory of gravity determines the universality of neutron star properties \cite{Vylet2024,Alwan2025}. However, not all alternative theories differ from GR; for instance, some models with non-linear gravity-matter coupling do not \cite{Sham2013}. To disentangle the roles of gravity and the equation of state (EoS) in neutron star structure, Ref. \cite{Eksi2014} used scalar curvature in GR as an indicator, arguing that neutron star structure constrains gravity more strongly than the EoS.
	
	In this work, we use an indicator similar to Ref.~\cite{Eksi2014} to describe the space-time structure in compact stars. We examine dimensionless central, surface, and volume-averaged curvature scalars (Ricci, Ricci tensor contraction, Kretschmann, Weyl), applying appropriate normalization. We pair these scalars with $I$ and $\Lambda$ to test their quasi-universal behavior. Using neutron star and quark star equations of state, we generate stellar structure solutions and related observables. We find that these curvature quantities connect to microscopic interior properties, prompting us to investigate the quasi-universality of related interior quantities with the same normalization with the curvature quantities. Lastly, we demonstrate how tidal deformability measurements constrain scalar curvatures and interior properties of a canonical mass neutron star.
	
	The paper is organized as follows. The next section outlines the methods used for calculating stellar structure and analyzing scalar curvature properties. Section~\ref{SECT:universalrelationdefinition} reviews established quasi-universal relations and details the computation of $I$ and $\Lambda$ in this work. Sect.~\ref{SECT:resultanddiscussion} presents our main findings and interprets their significance. Finally, Sect.~\ref{SECT:conclusion} summarizes the key outcomes and discusses broader implications. We use geometrized units throughout ($G=c=1$), unless specified otherwise.

	\section{Curvature scalars in spherical massive compact stars}
	
	In this paper, we consider two types of relativistic compact objects: neutron stars (NSs) and quark stars (QSs). NSs are thought to contain neutron-degenerate matter or exotic states and form at the end of massive stellar evolution. Their masses range from $1M_\odot$ to just under $3M_\odot$. QSs are more theoretical and modeled as objects of degenerate quark matter. They can have similar masses to NSs, but are usually smaller in radius due to being more compact. Both objects have extremely strong gravitational fields, making relativistic effects important for their internal and external structure. 
	
	\subsection{Canonical picture of a static, spherical relativistic star}
	As a dense, massive compact object, a NS and QS can be modeled as a relativistic, self-gravitating object. In general relativity (GR), the pressure and mass structure is calculated using the spherical interior solution of the Einstein field equation. The matter is assumed to be a perfect fluid, $T^{ab}=(\varepsilon+P)u^a u^b + P g^{ab}$. Considering the continuity fluid, i.e., $\nabla_\mu T^\mu = 0 $, we can solve for the pressure gradient to have the so-called Tolman-Oppenheimer-Volkoff (TOV) equation as \cite{Oppenheimer1939,Tolman1939}
	\begin{align}
		\label{eq:TOV}
		\frac{dP}{dr} &= -\frac{(\varepsilon + P)(m + 4\pi r^3 P)}{r^2(1 - 2m/r)},
	\end{align}
	with pressure, energy density, mass, and radial position as $P=P(r), \varepsilon=\varepsilon(r), m=m(r)$, and $r$, respectively. Meanwhile, the mass structure is obtained from the time-time component of the field equation, which is given by
	\begin{align}
		\label{eq:TOV-mass}
		\frac{dm}{dr} &= 4\pi r^2 \varepsilon.
	\end{align}
	
	The above equations can then be numerically solved by employing the EoS, denoted by $P=P(\varepsilon)$. For NSs, we used a well-known approach to represent several realistic equations of state, using piecewise polytropic EoS representations. Here, a set of parameters taken from Ref. \cite{Read2008}, which includes 3-point segments plus crust EoS. This model is used to represent 34 realistic EoSs with about $4\%$ of rms errors with respect to their corresponding tabular EoS. For QSs families, we use a color-flavor-locked EoS set from Ref. \cite{Flores2017} (contains 19 EoSs). This extends the traditional MIT-bag model and captures more variations in the parameters. 
	
\begin{figure}
		\centering
		\includegraphics[width=0.32\textwidth]{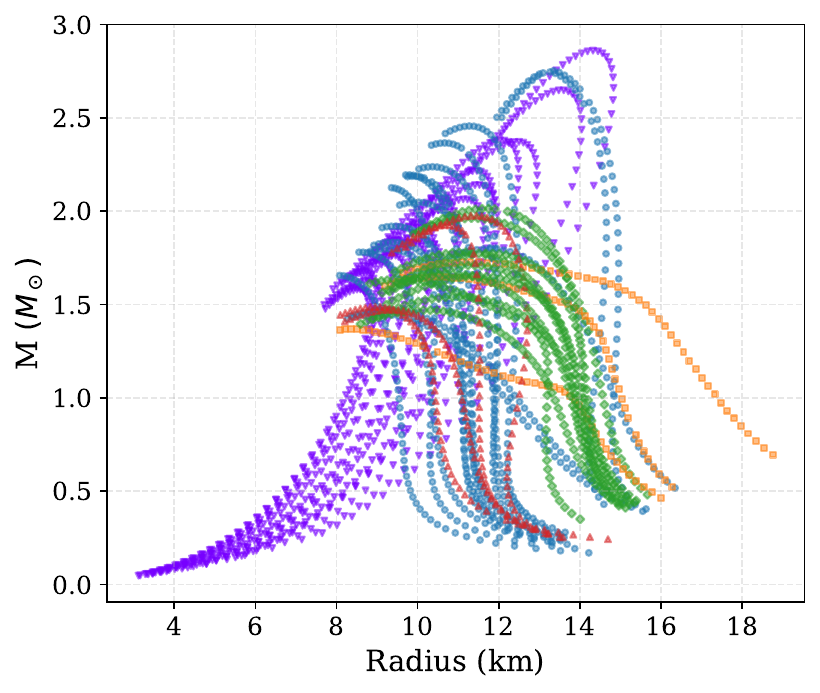}
		\hfill
		\includegraphics[width=0.32\textwidth]{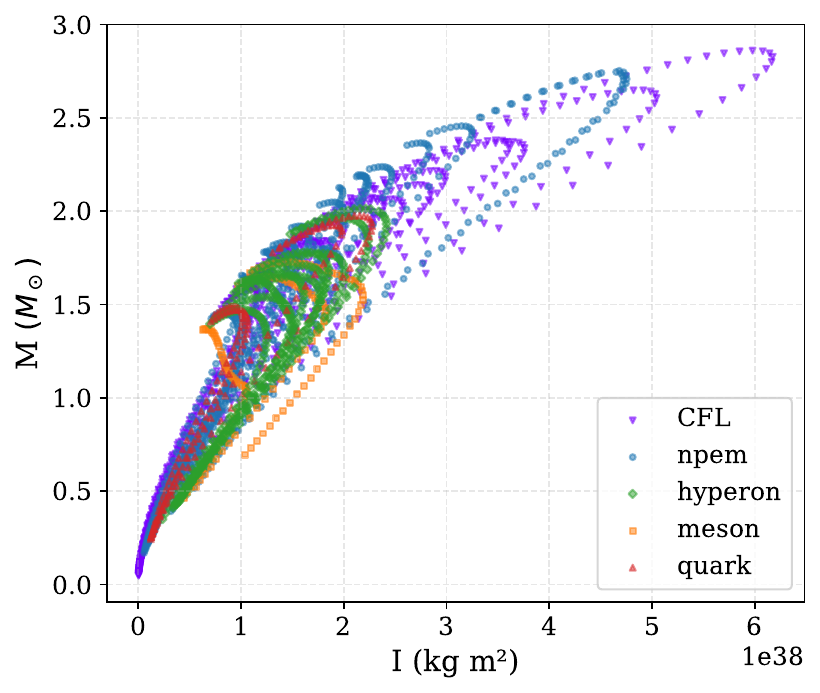}
		\hfill
		\includegraphics[width=0.32\textwidth]{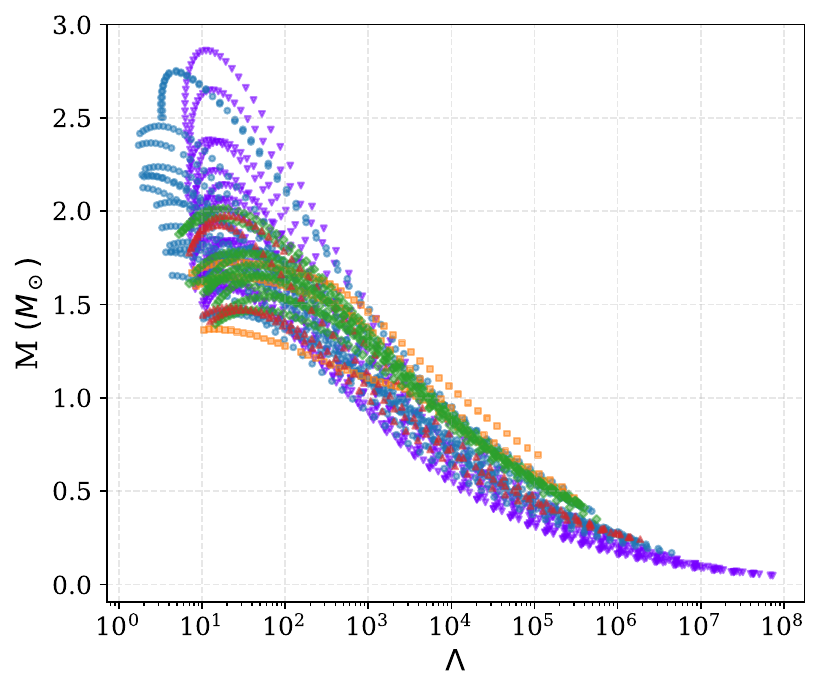}
		\caption{
			Mass–radius (left panel), mass–moment of inertia (middle panel), and mass–tidal deformability (right panel) relations are shown for the EoS sets used in this study. The models can be grouped as follows: (i) neutron stars with nucleonic (npem), hyperonic, mesonic, and quark matter compositions from Ref. \cite{Read2008}, and (ii) quark stars in the color–flavor–locked (CFL) phase from Ref. \cite{Flores2017}.
		}
		\label{fig:mass_radius_I_lambda}
	\end{figure}

	The mass-radius relations for both NS  and QS  EoS families can be seen in Fig. \ref{fig:mass_radius_I_lambda}. Both NS and QS EoS families in this set span masses below $3 M\odot$. However, their radius trends differ significantly. For NSs, lower-mass stars exhibit larger radii, while QS models show smaller radii for lower masses. This distinction arises from surface conditions: in the NS case, the crust (governed by the SLy4 EoS) contains heavy nuclei, and the density smoothly approaches zero as the pressure decreases to zero. This enables the star to have a thick, non-dense crust and, consequently, a large radius. In contrast, QS models have self-bound quark matter surfaces, resulting in non-zero density even as the pressure vanishes. This results in a QS maintaining a high-density structure for a low mass star.

	\subsection{Scalar curvatures in spherical stars}
	
	Several scalar quantities describe the curvature of spacetime, which is directly related to the gravitational tidal force through the Riemann tensor. In general relativity, a well-known example is the Ricci scalar. This is a trace component of the Riemann tensor and is directly connected to both the gravitational action and the field equations. Here, following Ref. \cite{Eksi2014}, we consider several curvature scalars inside a spherically symmetric mass distribution. These include the Ricci scalar ($\mathcal{R}$), the full contraction of the Ricci tensor ($\mathcal{J}$), the Kretschmann scalar ($\mathcal{K}$), and the full contraction of the Weyl tensor ($\mathcal{W}$). Please refer to Ref. \cite{Eksi2014}  for details. For simplicity, we omit "squared" notation on scalar symbols.

	The Ricci scalar indicates how local spacetime curvature differs from flat space around compact objects. It dictates the object's internal structure, stability, and gravitational properties.
	In terms of isotropic matter properties, the relevant expression can be derived from the Einstein field equation, which is given by
	\begin{align}
		\mathcal{R} = 8\pi(\varepsilon - 3P).\label{Eq:RicciScalar}
	\end{align}
	Based on above expression, it is clear that the Ricci scalar also directly represents the "trace anomaly", $\varepsilon-3P$, discussed in \cite{Fujimoto2022} which used as a signature of the conformality. 
	
	Similiarly, the full contraction of the Ricci tensor $ \mathcal{J} $ is defined by
	\begin{align}
		\mathcal{J} &\equiv R_{ab}R^{ab}\\ 
		\quad &=(8\pi)^2(\varepsilon^2 + 3P^2).\label{Eq:JScalar}
	\end{align}
	The Ricci tensor's contraction for a compact object encodes how volumes deform and expand or contract along geodesics. Its trace, the Ricci scalar, indicates the total matter-energy content. However, this scalar alone does not fully describe the gravitational field because it ignores directional stresses and fluxes. To analyze tidal effects, shape deformation, and gravitational curvature, the contraction of full Riemann tensor are required, which can be represented by the Kretschmann scalar and Weyl tensor.
	Note that both the Ricci tensor contraction and Ricci scalar become zero in vacuum regions and reach their maximum values at the center of the star.
	
	Then, the Kretchmann scalar $\mathcal{K}$ is defined by a full contraction of the Riemann tensor. Related to the Riemann tensor, this scalar is sensitive to the presence of singularities and provides a measure of the local curvature strength.
	\begin{align}
		\mathcal{K} &\equiv R^{abcd} R_{abcd},\\
		& = 64\pi^2(3\varepsilon^2 + 3P^2 + 2\varepsilon P) - \frac{128\pi m\varepsilon}{r^3} + \frac{48m^2}{r^6}.\label{Eq:KretScalar}
	\end{align}
	It is essential to note that the above expression comprises both matter terms and gravitational potential terms, rendering it non-vanishing in both the vacuum and within matter. In vacuum, only the last term remains, so $\mathcal{K}_{surf} = \frac{48m^2}{r^6}$. This expression is typically used to assess the presence of a "true" singularity—one that is invariant under coordinate changes—in the Schwarzschild metric, such as a black hole. The Kretchmann scalar captures the overall strength of gravity both inside the stellar interior and on the exterior.
	
	Lastly, the full-contraction of Weyl Tensor $\mathcal{W}$ is defined by 
	\begin{align}
		\mathcal{W} &\equiv \mathcal{C}^{abcd} \mathcal{C}_{abcd},\\
		&=\frac{4}{3}\left(\frac{6m}{r^3} - 8\pi\varepsilon\right)^2.\label{Eq:WeylFullScalar}
	\end{align}
	The Weyl tensor $\mathcal{C}_{abcd}$ is defined as the traceless part of Riemann tensor $R_{abcd}$. Similar to the Kretchmann scalar, the above expression also contains both matter and potential components. However, unlike the Kretchmann scalar, this  traceless part of the Riemann tensor isolates the measure of the distortion due to the tidal effects, a crucial aspect of compact object gravity. The value will approach zero at the stellar center, and be reduced to $\mathcal{W}_{vac} = \frac{48m^2}{r^6} = \mathcal{K}_{vac} $ towards the surface (vacuum).
	
	\section{Universal Relations: Compactness, $\Lambda$, and $I$}
	\label{SECT:universalrelationdefinition}
	
	The (quasi-)universal relation in compact stars discussed here pertains to cases where two or more correlated stellar quantities exhibit very low sensitivity to the equation of state (for the earliest example, please see the review by \cite{Doneva2017} for more historical discussions). As highlighted in \cite{Lattimer1989,Lattimer2001}, the normalized binding energy ($BE/M$) correlates with compactness ($C$) and is largely insensitive to the EoS, which can be approximated by $ BE/M \approx 0.6 C/(1 - C/2)$ (see the reference for details). In other words, $BE/M$ versus $C$ in neutron stars is nearly EoS-insensitive and can be described by a function $BE/M(C)$. Additionally, some astroseismology quantity pairs are also EoS-insensitive. Early examples include: $f$-mode frequency versus average density, $\omega_f (MR^{-3})$; $f$-mode damping time versus compactness, $\tau_f M^{-3} R^{4} (C)$; and, for the first $w$-mode, the EoS-insensitive functions $R \omega_w (C)$ and $M \tau_w^{-1} (C)$ \cite{Andersson1997,Andersson1996}.
	More recent studies have shown relations between compactness and other observable quantities, such as the moment of inertia $I$ and quadrupole $q$. These include $IM^{-1}R^2(C)$ and $q \equiv QMJ^{-2} = q(C)$ \cite{Lattimer2001,Bejger2002,Lattimer2005}. These two quantities were further investigated by the authors of \cite{Yagi2013,Yagi2013a}.
	
	We follow the $I-$Love notation from Refs. \cite{Yagi2013,Yagi2013a,Yagi2017} which relates the moment inertia with tidal Love number. The following discussion outlines the methods used to calculate moment inertia and tidal deformability in this study.
	
	\textit{Tidal deformability -- $\Lambda$.} To calculate the tidal deformability, we follow the method from Refs. \cite{Thorne1967,Hinderer2008,Hinderer2009,Krastev2018,Postnikov2010}, assuming a static, spherically symmetric star within a binary system influenced by the gravitational potential from its binary counterpart, $\mathcal{E}_{ij}$. In response, the stellar quadrupole moment $Q_{ij}$ deforms with a magnitude proportional to $\lambda_2$, a tidal deformability parameter related to the tidal Love number $k_2$. The explicit relation can be written as follows:
	\begin{align}
		Q_{ij} = -\lambda_2 \mathcal{E}_{ij},
	\end{align}
	where negative sign indicate that $Q_{ij}$ is opposing the tidal field $\mathcal{E}_{ij}$.
	
	The dimensionless tidal deformability $\Lambda$ then introduced by
	\begin{align}\label{EQ:lambdadef}
		\Lambda \equiv \lambda_2/M^5 = \frac{2}{3} k_2 / C^5.
	\end{align}
	
	The value of $k_2$ can be calculated by modeling a static, linear perturbation to the spherical metric \cite{Thorne1967}. the perturbation equations around a spherical metric (See Refs. \cite{Hinderer2008,Hinderer2009} for complete derivation). The exact value of $k_2$ is then can be
	calculated by solving the differential equation for $ y(r) $, which is given by \cite{Postnikov2010}
	\begin{align}
		\frac{dy}{dr} = \frac{-y^2 - yF - r^2 Q}{r},
	\end{align}
	where
	\begin{equation}
		F = \frac{1 - 4\pi r^2 (\varepsilon - P)}{1 - 2m/r},
	\end{equation}
	\begin{align}
		Q(r) &= 4\pi \left[ 5\varepsilon + 9p + \frac{\varepsilon + p}{\mathrm{d}P/\mathrm{d}\varepsilon} - \frac{6}{4 \pi r^2} \right] 
		e^{\lambda} \notag \\
		&\quad - \frac{4m^2}{r^4} \left[ 1 + \frac{4\pi r^3 p}{m} \right]^2 e^{2\lambda},
	\end{align}
	with $e^{\lambda} = \left[ 1 - \frac{2m}{r} \right]^{-1}$. Solving above differential equation simultaneously with the stellar structure using $y(r=0)=2$ as the initial condition, we will obtain the $y_R \equiv y(R)$ solution for then used to calculate
	\begin{align}
		k_2 &= \frac{8}{5}C^5(1-2C)^2[2 + 2C(y_R-1) - y_R] \\
		&\quad \times \{2C[6-3y_R + 3C(5y_R-8)] \\
		&\quad + 4C^3[13-11y_R + C(3y_R-2) + 2C^2(1+y_R)] \\
		&\quad + 3(1-2C)^2[2-y_R + 2C(y_R-1)]\ln(1-2C)\}^{-1},
	\end{align}
	with $C = M/R $. From this, the tidal deformability $\Lambda$ can be determined via Eq. (\ref{EQ:lambdadef}). For non-zero surface density stars such as QS, extra surface density term on $y$ (i.e., from Eq. (15) of Ref. \cite{Hinderer2009}) is added.
	
	\textit{Moment inertia -- $\bar{I} \equiv I/M^3$.} The moment of inertia, defined as angular momentum ($J$) divided by angular velocity ($\Omega$) $J/\Omega$, is calculated using the method from Refs. \cite{Hartle1967,Hu2023}.  The moment of inertia of a slow-rotating relativistic sphere can be calculated by solving the differential equation of \cite{Hartle1967,Hu2023,Dong2023,Wu2025}
	\begin{align}
		\frac{dI}{dr} = \frac{8}{3}\pi r^4(\varepsilon + P)\frac{1 - \frac{5I}{2r^3} + \frac{I^2}{r^6}}{1 - 2m/r}.
	\end{align}
	The above differential equation is done simultaneously within the stellar structure solver calculations with the initial condition of $I(r=0)=0$ and final $I=I(R)$.
	
	\section{Results and Discussions}
	\label{SECT:resultanddiscussion}
	
	Before further discussing the numerical results, we observe that the EoS-insensitive relations presented in Sect. \ref{SECT:universalrelationdefinition} are generally normalized by $M$ and/or $R$. These relations typically involve creating a dimensionless value by combining quantities, either in standard units (for example, $I/(MR^2)$) or in geometrized units (such as $I/M^3$). In line with these examples, the (quasi-)universal behaviors observed in our results can be achieved by normalizing quantities to transform them into dimensionless forms, either in standard or geometrized units. 
	
	In the standard $I-$Love relation \cite{Yagi2013}, $\Lambda$ is already defined as a dimensionless quantity. The moment of inertia is normalized by the cubic mass (i.e., $I/M^3$) to produce an EoS-insensitive relation \cite{Lau2009,Yagi2013a}. Alternatively, normalization by $I/MR^2$ also yields universality, as observed in earlier studies (e.g., \cite{Ravenhall1994}). In this study, our generated $I-\Lambda$ quantities are consistent with the standard $I-$Love universal relation. It is essential to note that the curvature quantities analyzed here are also normalized, rendering them dimensionless in geometric units. We maintain consistent units for all related quantities.
	
	All numerical calculations are carried out using a custom TOV-solver, which also performs computations for tidal deformability and moment of inertia, while simultaneously evaluating curvature scalars and other related quantities. We generate 50 log-equidistantly spaced stars for each EoS within $P_c \in [5,1300]$ MeV/fm$^{-3}$, for both NS and QS EoSs. 
	
	The output quantities from above calculations are then plotted into log-log scatter plot, while the trends are fitted into the 5-parameters log-space polynomial fit, which is defined by 
	
	\begin{align} \label{EQ:fittingfunction}
		\ln(Y) = \sum_{i=0}^{4} a_i [\ln(X)]^i,
	\end{align}
	with X and Y are NS quantities. The precision of the fitting function is then assessed by \textit{mean average residual error} (MARE) with the percentage is defined by
	\begin{align} \label{EQ:MAREdefinition}
		\mu_{(Y)}~(\%) = \text{average} \left| \frac{y - y_{\text{fit}}}{y} \right| \times 100\%,
	\end{align}
	where $y$ and $y_{\text{fit}}$ denote the actual and predicted values from the fit, averaged over all EoS for each stellar type (NS/QS). Another useful metric for assessing universality on a logarithmic scale is the percentage of stellar solutions that fall within a certain dex. A dex refers to a factor of ten difference on a logarithmic scale; for example, 2 dex corresponds to a factor of 100, and 3 dex to 1000. A threshold of $\pm 0.2$ dex is commonly used to assess whether a data point aligns with a trend line in logarithmic space. It is essential to note that the error metrics discussed above are strongly dependent on the chosen EoS set and the fitting function. For example, a larger EoS set (including both stiff and soft EoS) will exhibit a greater dispersion of values (see Ref. \cite{Saes2021}). The linear scale error also depends on the range covered by a quantity. Quantities spanning several orders of magnitude tend to have larger linear errors than those spanning only one or a few orders of magnitude. For consistency, we use the same EoS sets in all relations studied here and present both linear errors $\mu_{(Y)}$ and logarithmic errors $\mu_{(\ln Y)}$ for each relation. 
	
	\subsection{Standard $I - $Love$ - C$ relation}
\begin{figure}
		\centering
		\includegraphics[width=0.32\textwidth]{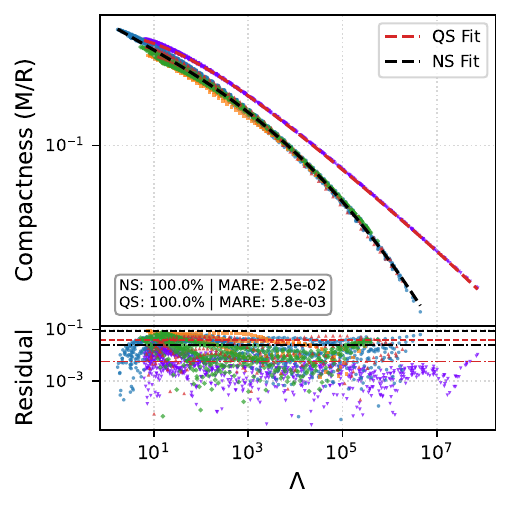}%
		\hfill
		\includegraphics[width=0.32\textwidth]{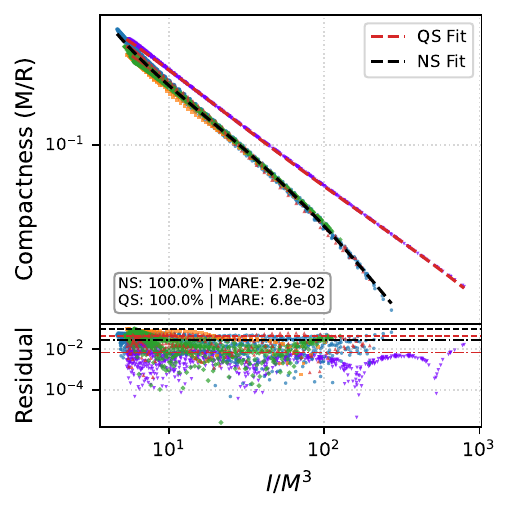}%
		\hfill
		\includegraphics[width=0.32\textwidth]{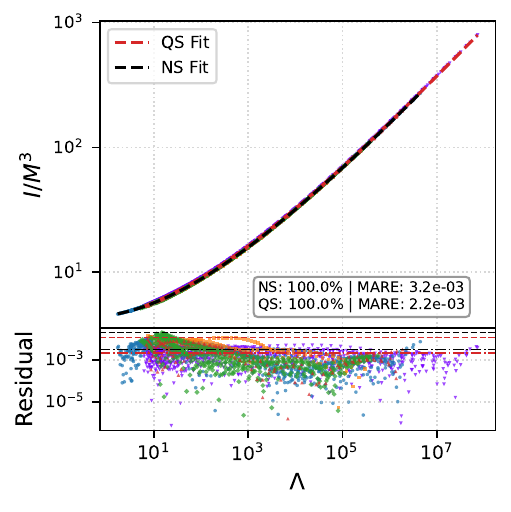}
		\caption{
			The universal $I$--Love--$C$ relations are reproduced by the stellar solution of the EoS set used in this study, with data points color-coded similarly to Fig. \ref{fig:mass_radius_I_lambda}. 
			The left and middle panels display the compactness $M/R$ as a function of the tidal deformability $\Lambda$ and the normalized moment of inertia $I/M^{3}$, respectively. The right panel shows the $I$--Love relation between $I/M^{3}$ and $\Lambda$. The percentages indicate the fraction of stellar solution data points within 0.2 dex of the fit, while MARE denotes the mean absolute residual error. The lower subpanels present the residuals relative to the fits for each relation. In the $I$--$\Lambda$ relation, neutron stars and quark stars overlap along a common trend. 
		}
		\label{fig:I_lambda_C}
	\end{figure}

	Here, we demonstrate the well-known universal relation between $I/M^3 - \Lambda - C$. We use the sets of equations of state investigated in this study as the baseline for the subsequent results. Three relations are shown: $C-\Lambda$, $C-I/M^3$, and $\Lambda - I/M^3$, as seen in Fig. \ref{fig:I_lambda_C}. The polynomial fit and their errors for these three relations are compiled in Tab. \ref{Tab:I_lambda_C_table}. 
	
	The $I/M^3-\Lambda$ relation is consistent with Ref. \cite{Yagi2013} (see also their Erratum). It shows almost no difference between NS and QS. The polynomial fit exhibits an extremely small error ($\mu < 1\%$). In contrast, when considering the compactness, the scaling relation diverges between NS and QS. This applies to both $C-\Lambda$ and $C-I/M^3$. This characteristic, where NS and QS show different relations, has also been discussed in \cite{Lugones2025} and is present in \cite{Staykov2016}.
	
	In the above examples, the MARE is no more than $3\%$ for NS and under $1\%$ for QS. All stellar points lie within 0.2 dex of the trendline. The standard $I/M^3-\Lambda$ relation shows $\mu<1\%$. These results demonstrate that the well-known $I-\text{Love}-C$ universal relation is also well-recovered using EoS sets and the method employed. The error values obtained here are used as a benchmark to assess other relations studied. In the following sub-sections, the main parts of this work will be presented.

	\begin{table}
		\centering
		\caption{Fitting coefficients of the relation (Eq. \ref{EQ:fittingfunction}) for compactness $C = M/R$ and moment of inertia $I/M^3$ as a function of either $\Lambda$ or $I/M^3$. $\mu_{(\ln Y)}$ and $\mu_{(Y)}$ denote the mean average residual error in log scale and linear scale, respectively.}
		\begin{tabular}{cccccccrr}
			\hline\hline
			\multicolumn{9}{c}{$Y = C = M/R$} \\ \hline
			\textbf{Type} & X & $a_0$ & $a_1$ & $a_2$ & $a_3$ & $a_4$ & $\mu_{(\ln Y)}$ (\%) & $\mu_{(Y)}$ (\%) \\
			\hline
			\textbf{NS} & $\Lambda$ & -1.019e+00 & -1.251e-01 & 1.738e-03 & -5.163e-04 & 7.454e-06 & 1.488718 & 2.488280 \\
			\textbf{NS} & $I/M^3$  & 8.204e-01 & -1.931e+00 & 5.618e-01 & -9.794e-02 & 5.706e-03 & 1.770502 & 2.871945 \\
			\textbf{QS} & $\Lambda$ & -1.155e+00 & -2.568e-02 & -3.309e-02 & 2.444e-03 & -6.997e-05 & 0.374481 & 0.578279 \\
			\textbf{QS} & $I/M^3$ & -5.809e-01 & -1.892e-01 & -1.461e-01 & 2.574e-02 & -1.565e-03 & 0.479337 & 0.681043 \\
			\hline\hline
			\multicolumn{9}{c}{$Y = I/M^3$} \\ \hline
			\textbf{Type} & X & $a_0$ & $a_1$ & $a_2$ & $a_3$ & $a_4$ & $\mu_{(\ln Y)}$ (\%) & $\mu_{(Y)}$ (\%) \\
			\hline
			\textbf{NS} & $\Lambda$ & 1.502e+00 & 4.980e-02 & 2.482e-02 & -9.085e-04 & 1.434e-05 & 0.156157 & 0.319225 \\
			\textbf{QS} & $\Lambda$ & 1.455e+00 & 8.760e-02 & 1.823e-02 & -4.567e-04 & 3.428e-06 & 0.097776 & 0.218162 \\
			\hline\hline
		\end{tabular}
		\label{Tab:I_lambda_C_table}
	\end{table}

	\subsection{Correlation with the Central Curvature}
	We discuss the universal relation of dimensionless $I$ and $\Lambda$ (which is already defined as dimensionless) towards scalar curvature evaluated at the stellar center. At this point, the curvature is pronounced and affected strongly by the dense core matter. In the stellar center, the Weyl scalar, a curvature invariant measuring tidal distortion, approaches $\mathcal{W} \rightarrow 0$ \cite{Eksi2014} and is affected by higher orders of density distribution \footnote{This consequently challenges our numerical and equation of state (EoS) setup, as we use piecewise polytropic functions to represent the EoS}. Therefore, in this part, we focus on calculating the central values of the squared Ricci scalar ($\mathcal{R}_c$), the squared Kretschmann scalar ($\mathcal{K}_c$), and the squared trace-free Ricci tensor ($\mathcal{J}_c$) to probe the scalar curvature in the stellar center. Due to the very dense environment in the stellar center, central quantities are dominated by the microscopic term, rather than the potential term [e.g., first term on Eq. (\ref{Eq:KretScalar})].
	
\begin{figure}[t]
		\centering
		\includegraphics[width=0.32\textwidth]{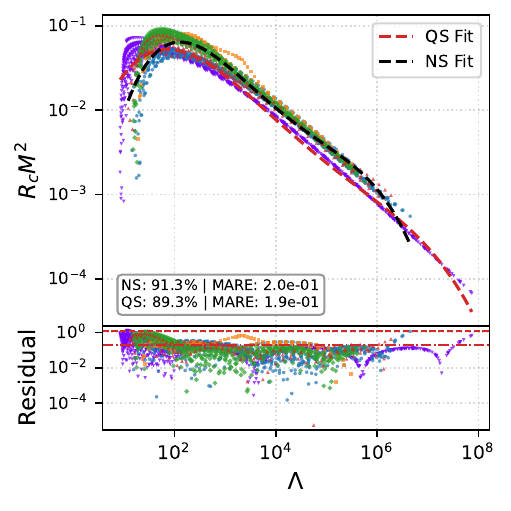}%
		\hfill
		\includegraphics[width=0.32\textwidth]{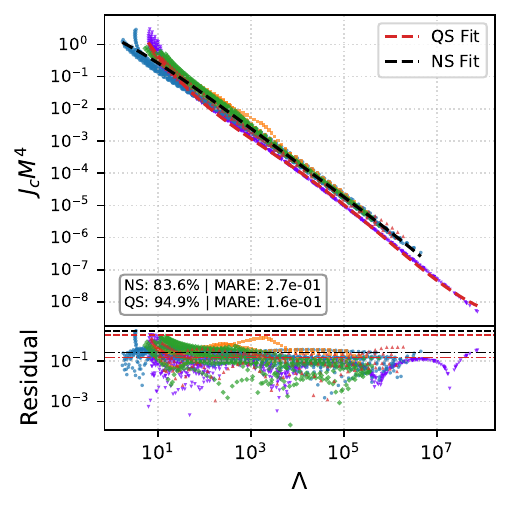}%
		\hfill
		\includegraphics[width=0.32\textwidth]{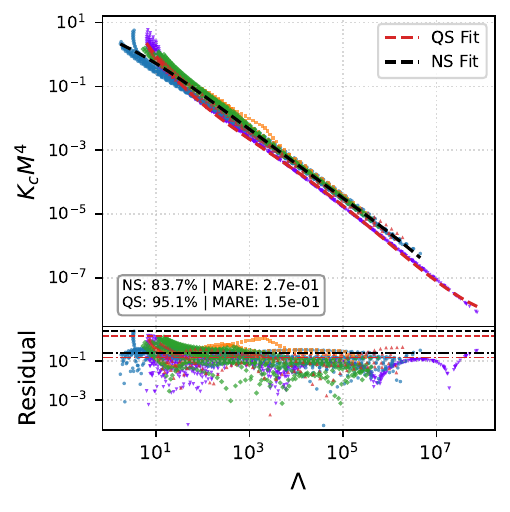}\\[1ex]
		\includegraphics[width=0.32\textwidth]{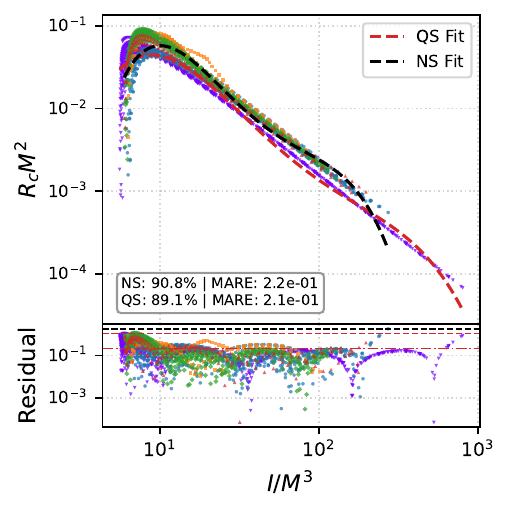}%
		\hfill
		\includegraphics[width=0.32\textwidth]{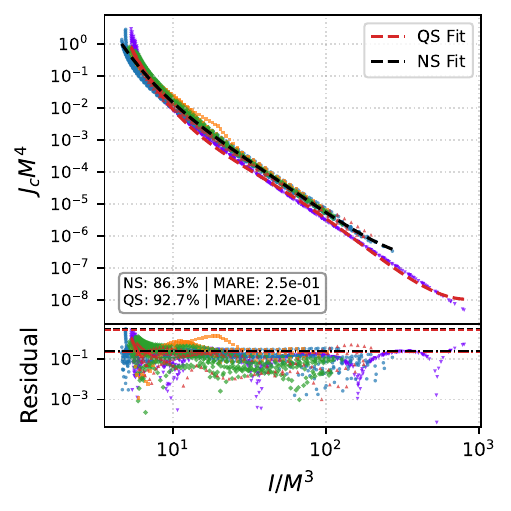}%
		\hfill
		\includegraphics[width=0.3\textwidth]{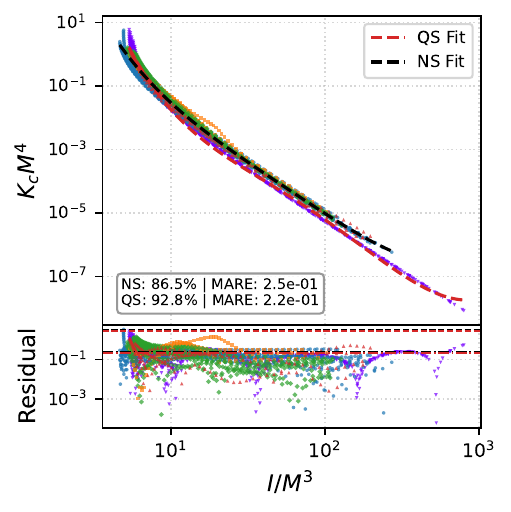}
		\caption{The universal central curvature $\mathcal{R}_c-\mathcal{J}_c-\mathcal{K}_c$ relations (normalized with $M$) obtained from the stellar solutions of the EoS set, with color categorization similar to Fig.~\ref{fig:mass_radius_I_lambda}. The upper panels show the relations with tidal deformability $\Lambda$, while the lower panels show the corresponding relations with $I/M^{3}$. 
			The percentages indicate the fraction of data points lying within 0.2 dex of the fit, MARE denotes the mean absolute residual error, and the lower subpanels present the residuals relative to the fits, similar to Fig.~\ref{fig:I_lambda_C}. Both neutron star (NS) and quark star (QS) sequences are represented by the polynomial fits on Eq. (\ref{EQ:fittingfunction}) with coefficients listed in Table  \ref{tab:fitting_coeff_NS_Lam}-\ref{tab:fitting_coeff_QS_I}.}
		\label{fig:central_curvature_relations}
	\end{figure}

	Our scaling relations of these quantities are summarized in Fig. \ref{fig:central_curvature_relations}. In almost all cases, the central curvature, defined as values of $\mathcal{R}_c$, $\mathcal{J}_c$, and $\mathcal{K}_c$, monotonically decreases as dimensionless $\Lambda$ and reduced moment of inertia ($I/M^3$) increase, except for $\mathcal{R}_c M^2$. The normalized central Ricci curvature ($\mathcal{R}_c M^2$), the dimensionless product of the Ricci scalar at the center and the square of stellar mass, shows a departure from linearity at higher mass (i.e., investigated at lower mass). Around the maximum mass, the curve shows a maximum, which limits the value to no more than $\mathcal{R}_c M^2 \lesssim 10^{-1}$. This critical characteristic is observed in both neutron stars (NS) and quark stars (QS) with a similar value. 
	
	For both $\Lambda$ and $I/M^3$ pairs, the variances—meaning the spread in the data points—are seemingly heteroscedastic, which means they change with respect to the values of the quantities. Notably, variances are more prominent at higher stellar mass, that is, at small $\Lambda$ or small $I/M^3$. At lower mass, the solutions converge tightly along the trendline. Meanwhile, we notice that there are oscillating errors in the residuals, which represent the differences between the observed and predicted values. We confirm that this is not a physical phenomenon, but rather one dominated by the underfitting of the polynomial function, which cannot fully capture the tail of the trend. However, this does not significantly affect the calculation of the MARE, as the dispersion of stellar points in the larger mass zones is more dominant in the calculation of the mean.
	
	The scaling relation of these quantities, based on the MARE and $\pm 0.2$ dex coverage, is relatively low compared to the standard $I-$Love$-C$ relation. However, it shows a moderately good approximation for central curvature quantities. The results show that more than $89\%$ of points lie within $\pm0.2$ dex, reaching approximately $95\%$. These central quantity relations display an almost inseparable trend between NS and QS, even though the relationship can be visually distinguished in both $\Lambda$ and $I/M^3$ relations. 
	
	\subsection{Correlation with the Surface Curvature}
\begin{figure} 
		\centering
		\includegraphics[width=0.48\textwidth]{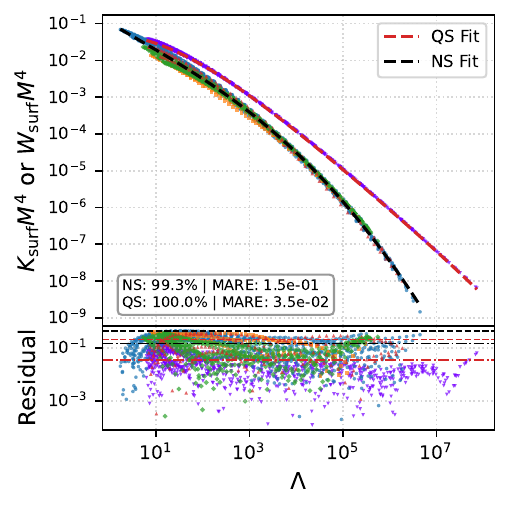}%
		\hfill
		\includegraphics[width=0.48\textwidth]{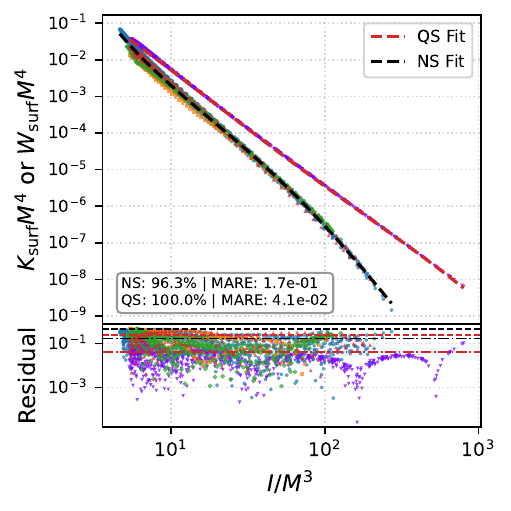}
		\caption{
			Similar to Fig.~\ref{fig:central_curvature_relations}, but for the normalized surface curvature relations with tidal deformability (left) and moment of inertia (right). At the stellar surface, only the Kretschmann scalar and the Weyl scalar do not vanish, and both yield identical values. Note that the overall trend shows very similar characteristics to the $C$-- $\Lambda$--$I$ relations shown in Fig.~\ref{fig:I_lambda_C}.
		}
		\label{FIG:surface_curvature_relations}
	\end{figure}

	It is interesting to evaluate the scalar curvature at the stellar surface, which serves as the transition zone between the stellar interior and the vacuum. At this point, only $\mathcal{K}$ and $\mathcal{W}$ remain, since $\mathcal{R}$ and $\mathcal{J}$ depend only on pressure and density. Surface quantities are defined from above—i.e., approached from the exterior—as $\mathcal{K}_s \equiv \mathcal{K}(R^+)$ and $\mathcal{W}_s \equiv \mathcal{W}(R^+)$ respectively. This leaves only the potential term in both $\mathcal{K}$ and $\mathcal{W}$, resulting in the identical expression $\mathcal{K}_s = \mathcal{W}_s = \frac{48M^2}{R^6}$. Universal relation plots for these quantities are shown in Fig. \ref{FIG:surface_curvature_relations}. 
	
	We observe that the trend closely resembles the $I/M^3-C$ or $\Lambda-C$ relation shown in Fig. \ref{fig:I_lambda_C}. Interestingly, the normalized relation directly represents the compactness, i.e., 
	\begin{align}
		\mathcal{K}_s M^4 = \mathcal{W}_s M^4  &=  \frac{48M^2}{R^6} M^4,\\ 
		& \propto C^6.
	\end{align}
	Therefore, from the opposite viewpoint, the compactness encodes the information about exterior gravitational quantities. Specifically, the quasi-universal behavior of quantities with respect to the compactness also represents their relation to the properties of the surface curvature.
	
	\subsection{Correlation with the Average Curvature}
	A quantity, such as curvature scalars or stellar structure microscopic quantities, is basically inhomogeneous inside a realistic stellar structure. We define a volume-averaged quantity to represent a single value of quantities for each star. Assuming a spherical star in flat spacetime, the volume-average quantity "$Q$" used in this study is defined by
	\begin{align}\label{Eq:averaging}
		\langle Q \rangle \equiv \frac{\int_0^R Q(r) \, dV}{\int_0^R dV} 
		\approx  \frac{3}{R^3} \int_0^R Q(r) \, r^2 \, dr.
	\end{align}
	with $dV=4\pi r^2 dr$ and $\int_0^R dV=\frac{4}{3} \pi R^3$ \footnote{For simplicity, we omit the relativistic correction to the volume, i.e., $dV = 4\pi r^2 \left( 1-\frac{2m(r)}{r}\right)^{-\frac{1}{2}} dr \approx 4\pi r^2 dr$. The simple form of $\langle Q \rangle$ will be useful to extend the universality to the derived quantities in a simple representation, which is discussed in Sect. \ref{Subsect:derivedquantities}.}. The $Q$ represents any stellar interior quantities which may vary throughout the stellar interior, such as curvature scalars and stellar interior properties. In this subsection, all curvature scalars are calculated to obtain their averaged values.

\begin{figure*}[t]
		\centering
		\includegraphics[width=0.24\textwidth]{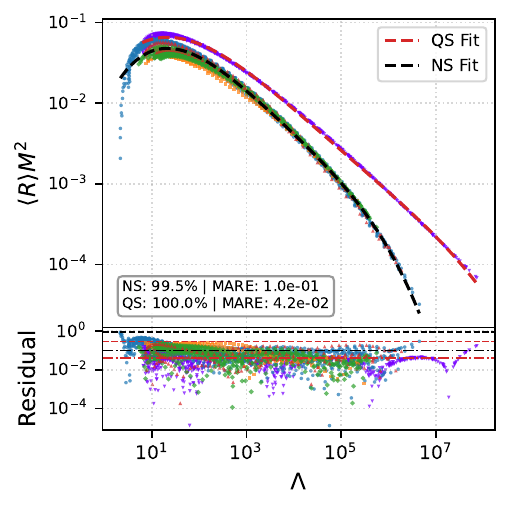}%
		\includegraphics[width=0.24\textwidth]{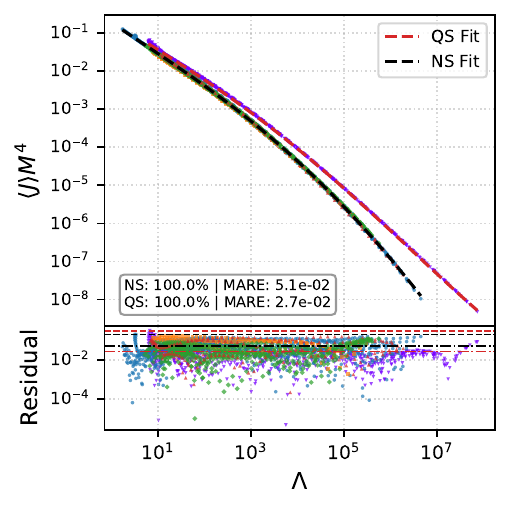}%
		\includegraphics[width=0.24\textwidth]{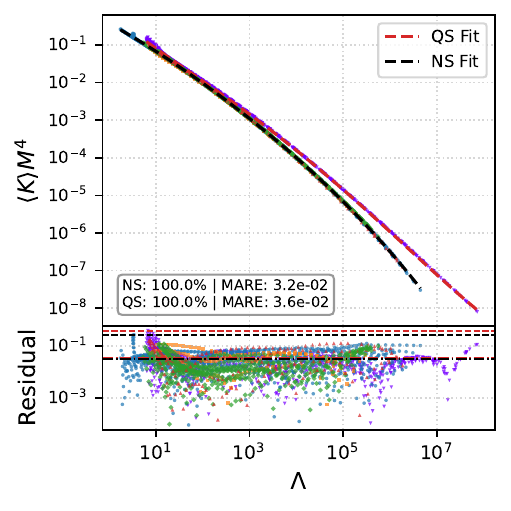}%
		\includegraphics[width=0.24\textwidth]{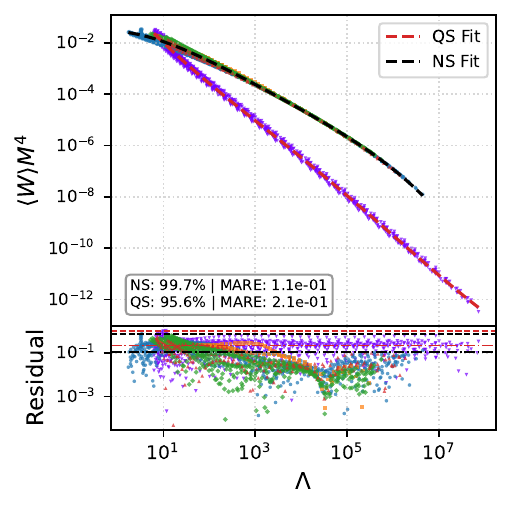}\\[1ex]
		
		\includegraphics[width=0.24\textwidth]{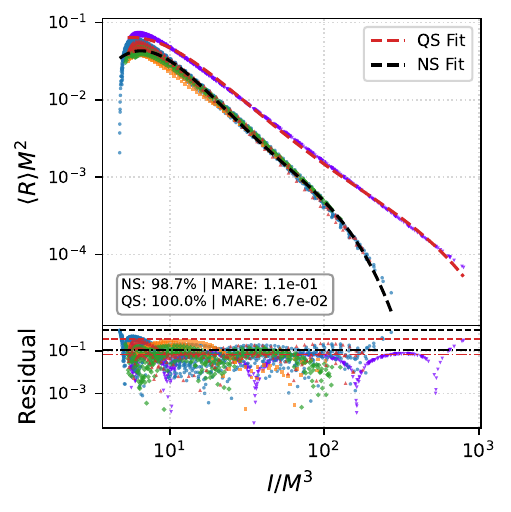}%
		\includegraphics[width=0.24\textwidth]{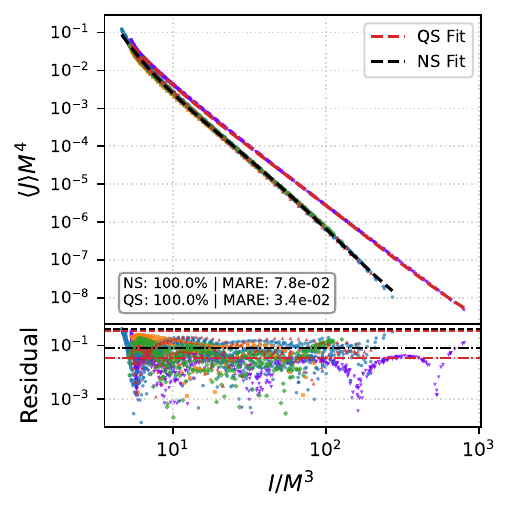}%
		\includegraphics[width=0.24\textwidth]{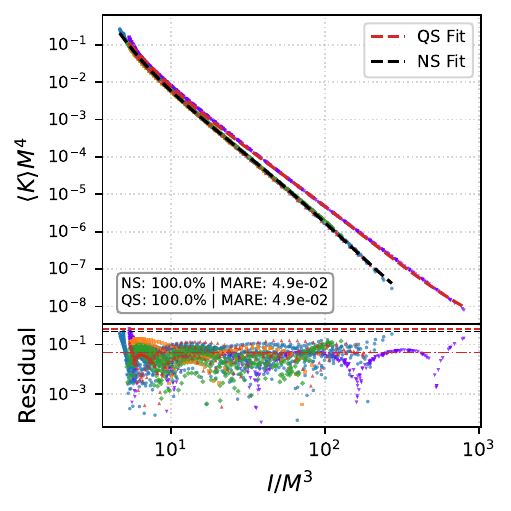}%
		\includegraphics[width=0.24\textwidth]{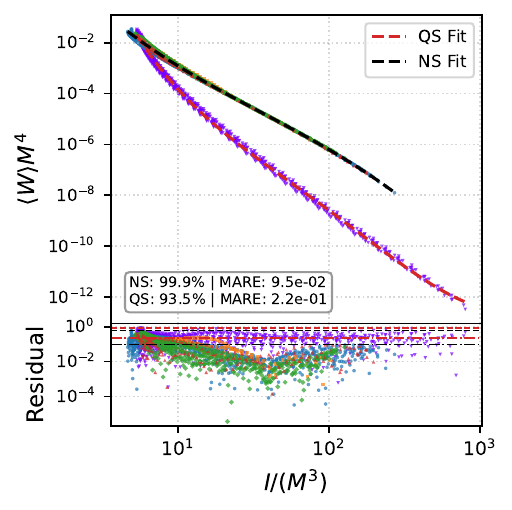}
		
		\caption{
			Similar to Fig.~\ref{fig:central_curvature_relations}, but for the volume-averaged curvature scalars ($\langle \mathcal{R} \rangle M^{2}$, $\langle \mathcal{J} \rangle M^{4}$, $\langle \mathcal{K} \rangle M^{4}$, and $\langle \mathcal{W} \rangle M^{4}$) relations with tidal deformability $\Lambda$ (upper) and moment of inertia $I/M^{3}$ (lower). 	
		}
		\label{fig:lambda_I_dimless_avg_4x2}
	\end{figure*}

	The above calculations are executed simultaneously within the stellar structure solver. The universal relation result is shown in Fig. \ref{fig:lambda_I_dimless_avg_4x2}. The universal relation trendlines showing a divergence similar to those of $I-C$  or $I-\Lambda$ relations between NS and QS. These behaviors are present in all averaged quantities. The differences between trends vary for different pairs of quantities. 
	
	The relation between averaged curvature quantities and $\Lambda-I/M^3$ generally shows smaller dispersion compared to the central (values at the core) and surface values, with uncertainties comparable to the $I$-Love-C relation. In particular, similar to the central curvature case, $\mathcal{\langle J \rangle}M^2$ and $\mathcal{\langle K \rangle}M^4$ show small dispersion, with less than $1\%$ average logarithmic residual error and $100\%$ of samples within $0.2$ dex. In contrast, $\mathcal{\langle R \rangle}M^2$ and $\mathcal{\langle W \rangle}M^4$ show more dispersed trends, with about $99\%-100\%$ of samples within $0.2$ dex, except for $\mathcal{\langle W \rangle}M^4$ in the QS case, where roughly $93\%-96\%$ are within $0.2$ dex.
	The turning point of the Ricci scalar is also present in the averaged value. For this case, the QS trend is slightly above the NS one. The maxima of $\mathcal{\langle R \rangle}M^2$ are similar with the $\mathcal{R}_c M^2$ case, with $\lesssim 10^{-1}$.

	\subsection{Correlation with other derived quantities} \label{Subsect:derivedquantities}
\begin{figure*}[t]
		\centering
		\includegraphics[width=0.32\textwidth]{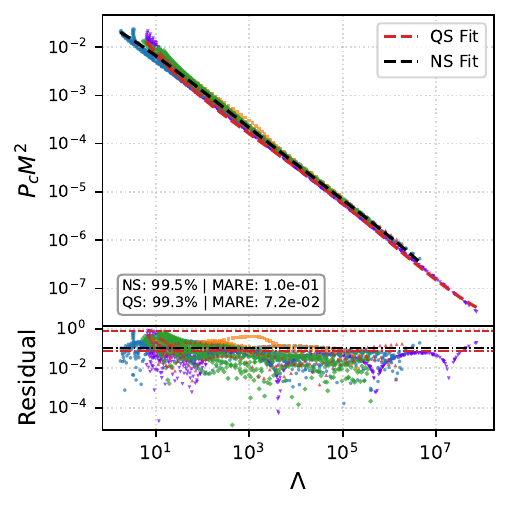}%
		\includegraphics[width=0.32\textwidth]{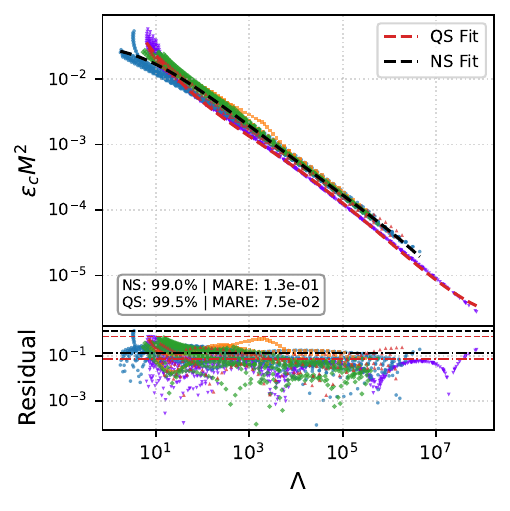}%
		\includegraphics[width=0.32\textwidth]{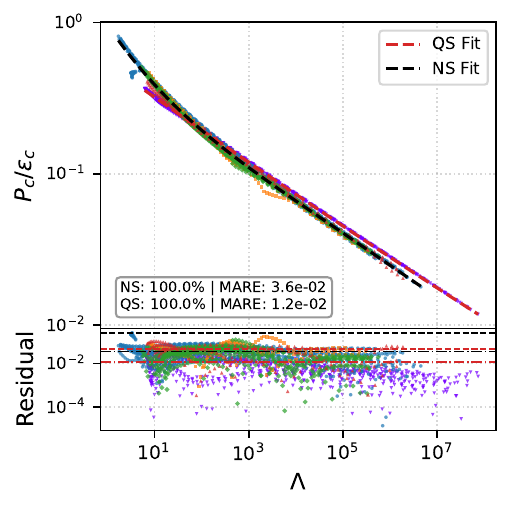}\\[1ex]
		\includegraphics[width=0.32\textwidth]{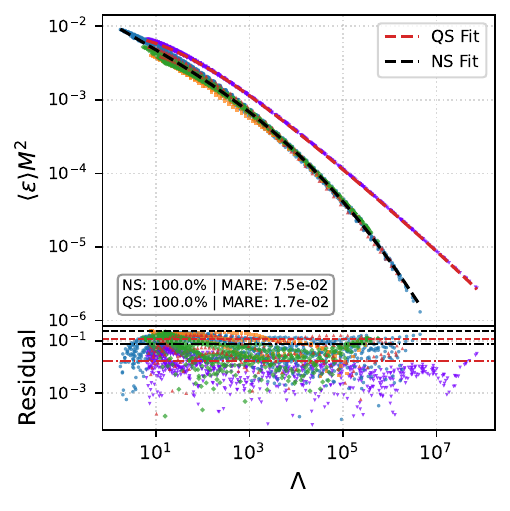}%
		\includegraphics[width=0.32\textwidth]{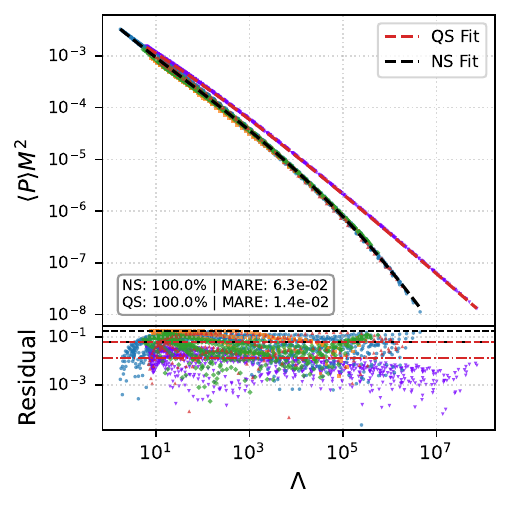}%
		\includegraphics[width=0.32\textwidth]{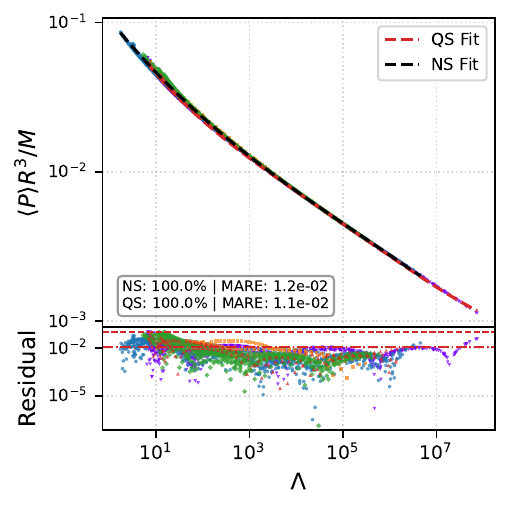}
		\caption{
			This figure is similar to Fig.~\ref{fig:central_curvature_relations}, but shows normalized pressure and energy density. The upper panels display the central quantities ($P_{c}/M^{2}$, $\varepsilon_{c} M^{2}$, and $P_{c}/\varepsilon_{c}$). The lower panels show volume-averaged quantities ($\langle \varepsilon \rangle/M^{2}$, $\langle P \rangle/M^{2}$, and $\langle P \rangle R^{3}/M$), all as functions of tidal deformability $\Lambda$.
		}
		\label{fig:lambda_vs_quantities}
	\end{figure*}

\begin{figure*}[t]
		\centering
		\includegraphics[width=0.32\textwidth]{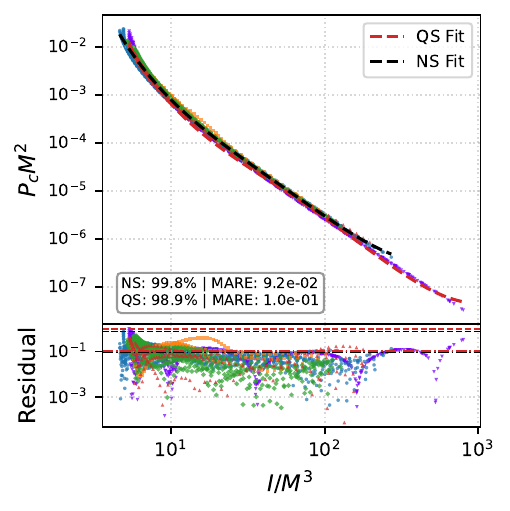}%
		\includegraphics[width=0.32\textwidth]{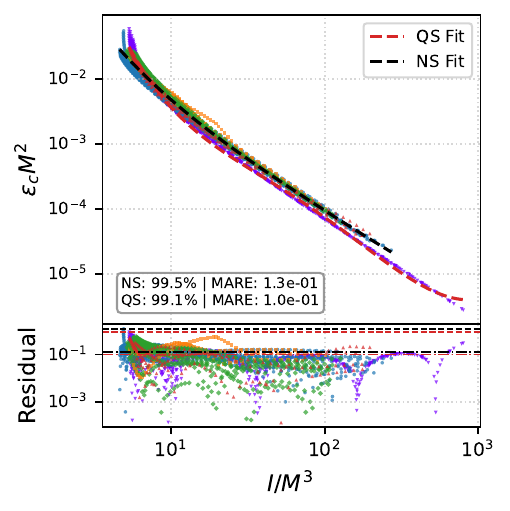}%
		\includegraphics[width=0.32\textwidth]{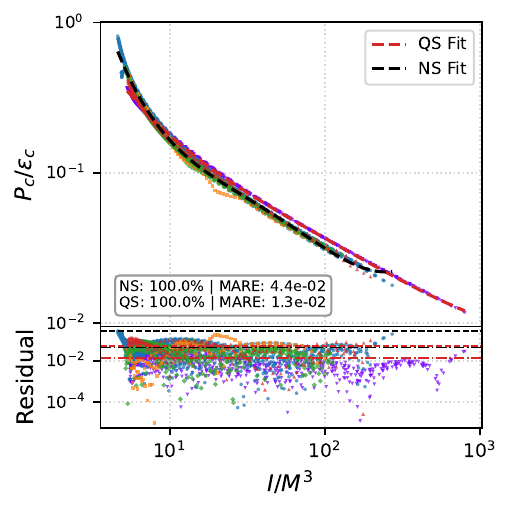}\\[1ex]
		\includegraphics[width=0.32\textwidth]{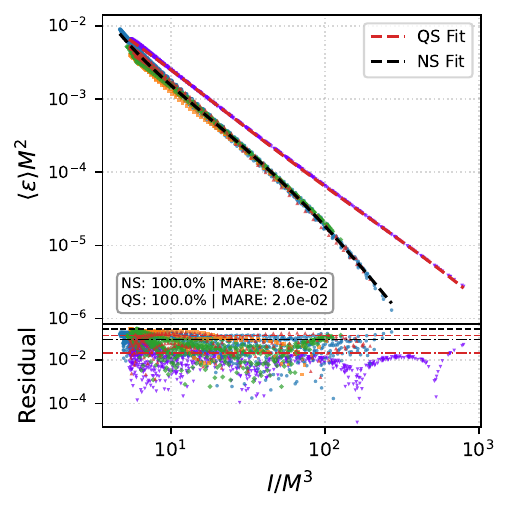}%
		\includegraphics[width=0.32\textwidth]{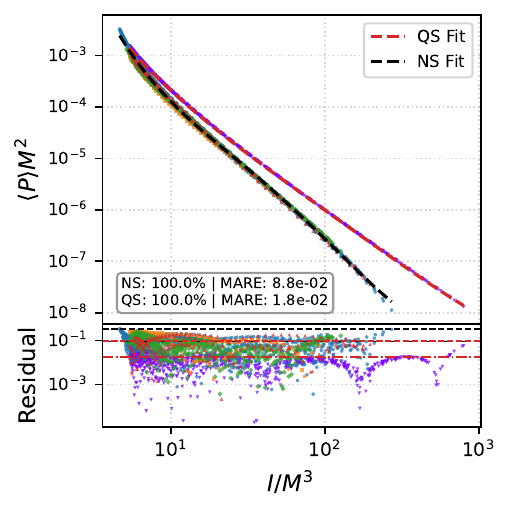}%
		\includegraphics[width=0.32\textwidth]{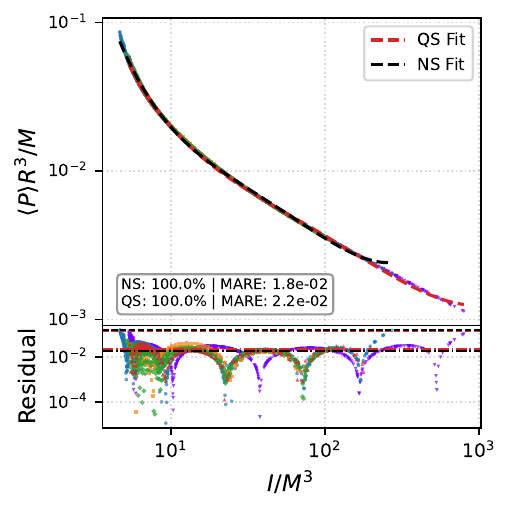}
		\caption{
			Similar to Fig. \ref{fig:lambda_vs_quantities}, but as the function of moment inertia $I/M^3$. 
		}
		\label{fig:I_dim_vs_quantities}
	\end{figure*}

	Generally, Eqs. (\ref{Eq:RicciScalar})--(\ref{Eq:WeylFullScalar}) show that curvature scalars are expressed in terms of matter quantities (such as pressure and energy density) as well as potential quantities (such as compactness, or mass-radius related terms). If these normalized curvature scalars show equation-of-state (EoS) insensitivity with respect to $I/M^3$ and $\Lambda$, then their similarly normalized constituents should also exhibit EoS-insensitive behavior. In this subsection, we further investigate the interior terms that appear in curvature scalars expression. Specifically, we focus on the central and average values of pressure and energy density, as well as their various combinations, to form new universal relations. We examine several simple combinations as examples, some of which reduce to known universal relations from the literature. The following quantities demonstrate strong universal relations with $I/M^3$ and $\Lambda$. The corresponding plots are shown in Figs. \ref{fig:lambda_vs_quantities} and \ref{fig:I_dim_vs_quantities}.
	\begin{enumerate}
		\renewcommand{\theenumi}{\alph{enumi}} 
		\renewcommand{\labelenumi}{(\theenumi)} 
		\item $\varepsilon_c M^2$ -- The energy density evaluated at the stellar center is normalized by the squared stellar mass. This term is extracted from the relation $\mathcal{R}_c \propto \varepsilon_c$. It is shown that the quantity $\varepsilon_c M^2$ decreases monotonically as both $I/M^3$ and $\Lambda$ increase. Both parameters show significant errors at higher masses and central energy densities. The trends for neutron stars (NSs) and quark stars (QSs) nearly overlap in both the $\Lambda$ and $I/M^3$ relations. 
		
		\item  $P_c M^2$ -- This quantity exhibits an analogous relationship to the $\varepsilon_c M^2$ case. A decreasing trend is also observed with respect to both $I/M^3$ and $\Lambda$, and larger errors are found for stars of greater mass. The trends for NS and QS in this case are also nearly overlapping.
		
		\item $P_c/\varepsilon_c$ -- A ratio between the two quantities above. This quantity represents the average EoS stiffness \cite{Saes2021} and related to average speed of sound via $\langle c_s^2 \rangle = \frac{1}{\varepsilon_c} \int_0^{\varepsilon_c} \frac{\text{d}P_c}{\text{d}\varepsilon_c} \text{d}\varepsilon$ as explained in Ref. \cite{Saes2024}. This quantity shows a small error relation with $I/M^3$ and $\Lambda$ compared with $\varepsilon_c M^2$ and $P_c M^2$ (see Table 	\ref{tab:fitting_coeff_NS_Lam}-\ref{tab:fitting_coeff_QS_I}).
		
		\item $\langle \varepsilon \rangle M^2$ -- The volume-averaged energy density is defined in Eq. (\ref{Eq:averaging}). The trend closely matches the $I-\Lambda-C$ relation, including its error behavior. This demonstrates that the quantity is tightly linked to compactness. Indeed, $\langle \varepsilon \rangle M^2 \propto (M/R^3)(M^2) \propto C^3$.

		\item $\langle P \rangle M^2$ -- Like (d) for pressure, this also shows a trend similar to $I-\Lambda-C$, though not an exact match. This suggests a possible correlation between $\langle P \rangle M^2$ and compactness. The average pressure connects to gravitational binding energy $W$ through the scalar virial theorem in hydrostatic equilibrium; specifically, $\langle P \rangle = -\frac{W}{3V}$, or roughly $\langle P \rangle \sim \frac{GM^2}{R^4}$. Thus, $\langle P \rangle M^2$ relates to compactness as $\langle P \rangle M^2 \sim C^4$. The gravitational binding energy approximates the baryonic binding energy in static, cold neutron stars, so this value also ties to the universality in binding energy, such as $BE/M$. A comparable relation with kinetic-to-gravitational binding energy $(T/W)$ appears in Ref. \cite{Krueger2023}.
		
		\item $\langle P \rangle R^3/M$ -- This quantity is calculated from the ratio of average pressure to average density, i.e., $\frac{\langle P \rangle  M^2}{\langle \varepsilon \rangle M^2} = \frac{\langle P \rangle}{M/\frac{4}{3} \pi R^3} \propto \langle P \rangle \frac{R^3}{M}$. The trend line for the relation is highly EoS-insensitive. It has small residual errors with the trend line also shows no gap between NS and QS. This is similar to the $I-\Lambda$ relation. Intuitively, this relation is connected to $\langle P \rangle \sim \frac{GM^2}{R^4}$, as explained in (e). This yields $\langle P \rangle \frac{R^3}{M} \sim C$. However, the universality result (Fig. \ref{fig:lambda_vs_quantities} and \ref{fig:I_dim_vs_quantities}) shows a very different pattern with the $I-\Lambda-C$ relation. Here, the NS and QS trends are tightly overlapping.
	\end{enumerate}

	The scaling relations associated with central EoS values have been extensively studied in Ref. \cite{Cai2025}. They performed a perturbative analysis on the reduced TOV equation, scaling quantities with their central values to make the TOV equation dimensionless. In the Newtonian approximation, they show that $P_c \sim M^2/R^4$. The relativistic version can be expanded in terms of compactness using a power-law form. With a similar method, they found that $\langle c_s^2\rangle$, directly connected to $P_c/\varepsilon_c$, is closely related to compactness. These two relations from the literature match the properties of our derived central universal relations, i.e., quantities (a)--(c).
	
	In contrast to the central relations, the averaged quantities show a stronger EoS-insensitivity than central quantities (except for $P_c/\varepsilon_c$). Intuitively, the averaged pressure and energy quantities are more closely related to the structural energy inside a star, which is also linked to the compactness. For instance, in the above examples, the $\langle P \rangle R^3/M - I/M^3$ relation is highly EoS-insensitive. This is evident as the trendline cannot accurately capture the trend, resulting in oscillating residuals for both NS and QS in a wide range of $I/M^3$ (similar to the Runge phenomenon). This observation further highlights the limitation of using residuals alone to assess EoS insensitivity and motivates the exploration of alternative metrics.
	
\begin{figure*}[t]
		\centering
		\includegraphics[width=0.24\textwidth]{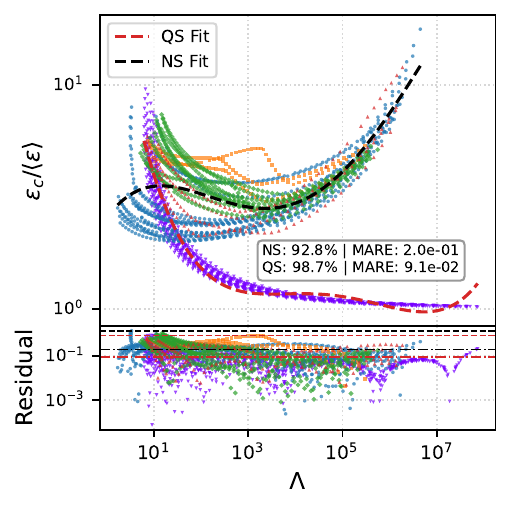}%
		\includegraphics[width=0.24\textwidth]{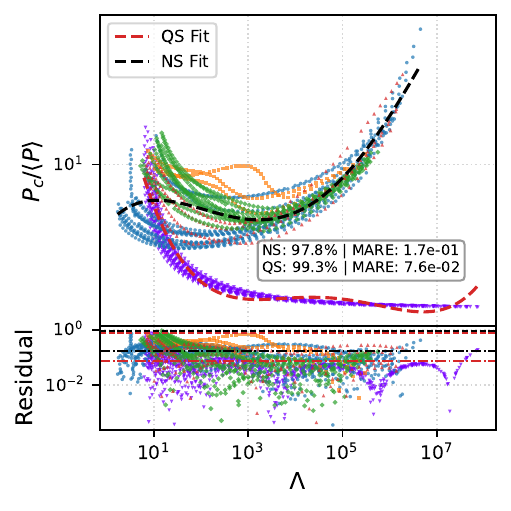}%
		\includegraphics[width=0.24\textwidth]{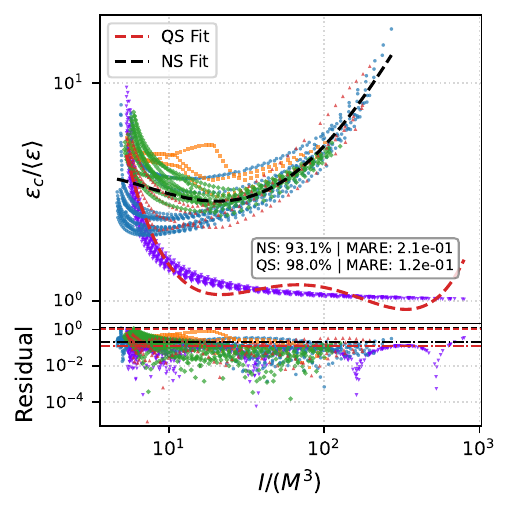}%
		\includegraphics[width=0.24\textwidth]{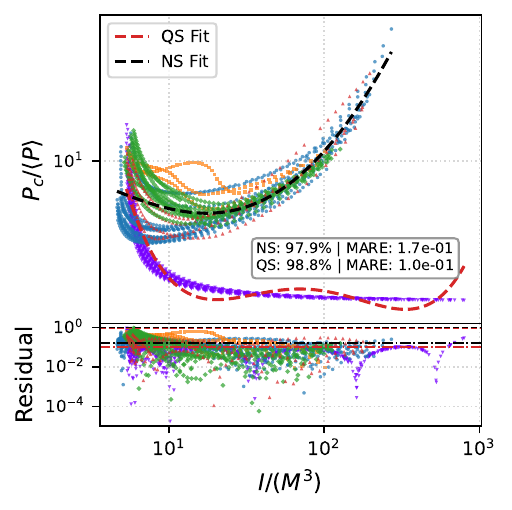}
		\caption{
			Similar to Fig.~\ref{fig:central_curvature_relations}, but for the relation of the central-to-average pressure and energy density ratio as a function of the tidal deformability $\Lambda$ and the moment of inertia $I/M^{3}$. The oscillatory behavior in the residuals arises because the polynomial fitting parameterization used here is unable to capture the trend precisely.
		}
		\label{fig:lambda_I_epsilon_pressure}
	\end{figure*}

	These examples, while simple, are intended for demonstration and naturally lead to a broader question. In principle, other combinations of the aforementioned quantities may also exhibit similar EoS insensitivity and, consequently, universality. However, it is important to note that some combinations do not show universal relational behavior, for instance, $\varepsilon_c/\langle \varepsilon \rangle$ and $P_c/\langle P \rangle$, as depicted in Fig. \ref{fig:lambda_I_epsilon_pressure}. This observation suggests the need for a more systematic investigation to assess the universality of other quantity combinations.
	
	Expanding on these patterns, we find that the universality exhibited by the derived central and average quantities closely mirrors that observed in curvature quantities, except for $\langle P \rangle/\langle \varepsilon\rangle$. For instance, the universal relation of central derived quantities, such as $P_c M^2$, $\varepsilon_c M^2$, and $P_c/\varepsilon_c$, follows a similar pattern to that of central curvatures: they are negatively correlated with heteroskedastic uncertainties, and the NS and QS curves nearly overlap. Meanwhile, average values such as $\langle \varepsilon \rangle M^2$ and $\langle P \rangle M^2$ show compactness-like trends, akin to averaged curvature quantities. Remarkably, the ratio between them, $\frac{\langle P \rangle  M^2}{\langle \varepsilon \rangle M^2}$ (or $\langle P \rangle R^3/M$), exhibits very small errors with no gaps between NS and QS trends, reinforcing the close connection to universality.
	
	\subsection{Bounds from multimessenger observations}
	The universal relations are shown to be useful in constraining neutron star properties through multiple observations \cite{Landry2018,Kumar2019,Godzieba2020}. In this section, we demonstrate how measurements of tidal deformability and, implicitly, the moment of inertia constrain the curvature and interior properties of a neutron star.
	
	Building upon these universal relations, we use the fitted "quantity" $Q-\Lambda$ relations from Table \ref{tab:fitting_coeff_NS_Lam}--\ref{tab:fitting_coeff_QS_I}. The observed value of $\Lambda$ serves as input to estimate curvature quantities and other derived properties. For example, we estimate the curvature and interior properties of canonical mass neutron stars using measured tidal deformability data from Ref. \cite{Abbott2018}, which reports $\Lambda_{1.4} = 190^{+390}_{-120}$ at a $90\%$ CI. Additionally, Ref. \cite{Kumar2019} obtained a more precise value by combining tidal deformability from GW170817 data and moment of inertia measured from low-mass X-ray binary observations via the $I/M^3-\Lambda$ relation, resulting in $\Lambda_{1.4} = 196^{+92}_{-63}$ at a $90\%$ CI. 
	
	The resulting predicted quantities for canonical-mass neutron stars, derived through the method described above, are presented in Table \ref{Tab:DeducedQTYS}. Both the curvature scalars and other related properties are constrained by fitted functions derived from the relation for each neutron star EoS. It is important to note that the errors are calculated solely from the $\Lambda_{1.4}$ measurements, without incorporating uncertainties from residual error.
	
	\begin{table*}[htbp]
		\centering
		\caption{Deduced of quantities for $M = 1.4\,M_\odot$ neutron star from GW170817 \cite{Abbott2018} ($\Lambda_{1.4} = 190^{+390}_{-120}$) and GW170817+LMXB \cite{Kumar2019} ($ \Lambda_{1.4} = 196^{+92}_{-63}$). $\exp(Y)$ are dimensionless in geometrical units as defined in Table \ref{tab:fitting_coeff_NS_Lam}-\ref{tab:fitting_coeff_QS_I}, while the 'Value' column shows deduced values of corresponding quantity with their unit on the rightmost column. These values are determined assuming the neutron star equation of state.}
		\label{Tab:DeducedQTYS}
		\begin{tabular}{lcccccr}
			\hline\hline
			\multirow{2}{*}{Quantity} & \multicolumn{2}{c}{GW170817} & \multicolumn{2}{c}{GW170817+LMXB} & \multirow{2}{*}{Unit} \\
			\cline{2-3} \cline{4-5}
			& $\exp(Y)$ & Value & $\exp(Y)$ & Value & \\ \hline
			$\mathcal{R}_{\mathrm{c}}$ & $0.0626^{[a]}_{-0.0174}$ & $1.4688^{+0.87^{[a]}}_{-0.4078}$ & $0.0623^{+0.0020}_{-0.0048}$ & $1.4623^{+0.0461}_{-0.1124}$ & $\mathrm{m}^{-2} \times 10^{-8}$ \\
			$\mathcal{K}_{\mathrm{c}}$ & $0.0274^{+0.0510}_{-0.0192}$ & $1.5085^{+2.8059}_{-1.0555}$ & $0.0265^{+0.0135}_{-0.0090}$ & $1.4592^{+0.7447}_{-0.4935}$ & $\mathrm{m}^{-4} \times 10^{-15}$ \\
			$\mathcal{J}_{\mathrm{c}}$ & $0.0142^{+0.0256}_{-0.0098}$ & $7.8149^{+14.0933}_{-5.4047}$ & $0.0137^{+0.0068}_{-0.0046}$ & $7.5651^{+3.7632}_{-2.5130}$ & $\mathrm{m}^{-4} \times 10^{-16}$ \\
			$\langle \mathcal{R} \rangle$ & $0.0298^{+0.0112}_{-0.0114}$ & $6.9988^{+2.6180}_{-2.6690}$ & $0.0295^{+0.0044}_{-0.0042}$ & $6.9162^{+1.0426}_{-0.9930}$ & $\mathrm{m}^{-2} \times 10^{-9}$ \\
			$\langle \mathcal{K} \rangle$ & $0.0054^{+0.0078}_{-0.0035}$ & $2.9549^{+4.2976}_{-1.9225}$ & $0.0052^{+0.0022}_{-0.0016}$ & $2.8716^{+1.2196}_{-0.8628}$ & $\mathrm{m}^{-4} \times 10^{-16}$ \\
			$\langle \mathcal{J} \rangle$ & $0.0023^{+0.0033}_{-0.0015}$ & $1.2670^{+1.7946}_{-0.8173}$ & $0.0022^{+0.0009}_{-0.0007}$ & $1.2319^{+0.5121}_{-0.3649}$ & $\mathrm{m}^{-4} \times 10^{-16}$ \\
			$\langle \mathcal{W} \rangle$ & $0.0011^{+0.0015}_{-0.0007}$ & $6.1676^{+8.4864}_{-3.9523}$ & $0.0011^{+0.0004}_{-0.0003}$ & $5.9983^{+2.4603}_{-1.7637}$ & $\mathrm{m}^{-4} \times 10^{-17}$ \\
			$\mathcal{W}_{\mathrm{surf}}$ & $0.0018^{+0.0024}_{-0.0012}$ & $1.0019^{+1.3217}_{-0.6383}$ & $0.0018^{+0.0007}_{-0.0005}$ & $0.9751^{+0.3867}_{-0.2816}$ & $\mathrm{m}^{-4} \times 10^{-16}$ \\
			$\mathcal{K}_{\mathrm{surf}}$ & $0.0018^{+0.0024}_{-0.0012}$ & $1.0019^{+1.3217}_{-0.6383}$ & $0.0018^{+0.0007}_{-0.0005}$ & $0.9751^{+0.3867}_{-0.2816}$ & $\mathrm{m}^{-4} \times 10^{-16}$ \\
			\hline 
			$\varepsilon_{\mathrm{c}}$ & $0.0046^{+0.0029}_{-0.0020}$ & $814.8607^{+512.3683}_{-356.3830}$ & $0.0045^{+0.0010}_{-0.0008}$ & $802.1846^{+171.4175}_{-142.7433}$ & $\mathrm{MeV}/\mathrm{fm}^{3}$ \\
			$P_{\mathrm{c}}$ & $7.5476^{+8.4813}_{-4.3174} \times 10^{-4}$ & $134.1344^{+150.7294}_{-76.7289}$ & $7.3716^{+2.5175}_{-1.8691} \times 10^{-4}$ & $131.0066^{+44.7408}_{-33.2178}$ & $\mathrm{MeV}/\mathrm{fm}^{3}$ \\
			$\frac{P_{\mathrm{c}}}{\varepsilon_{\mathrm{c}}}$ & $0.1644^{+0.0500}_{-0.0394}$ & $0.1644^{+0.0500}_{-0.0394}$ & $0.1631^{+0.0172}_{-0.0150}$ & $0.1631^{+0.0172}_{-0.0150}$ & $-$ \\
			$\langle \varepsilon \rangle$ & $0.0015^{+0.0008}_{-0.0006}$ & $261.2444^{+136.6131}_{-103.8790}$ & $0.0015^{+0.0003}_{-0.0002}$ & $257.7262^{+46.8461}_{-40.3749}$ & $\mathrm{MeV}/\mathrm{fm}^{3}$ \\
			$\langle P \rangle^{[b]}$ & $1.1963^{+1.2043}_{-0.6568} \times 10^{-4}$ & $21.2602^{+21.4021}_{-11.6732}$ & $1.1704^{+0.3657}_{-0.2788} \times 10^{-4}$ & $20.8001^{+6.4991}_{-4.9553}$ & $\mathrm{MeV}/\mathrm{fm}^{3}$ \\
			\hline\hline
		\end{tabular}
		\footnotetext{Because of the wide uncertainty around the $R_c M^2 - \Lambda$ peak, the upper limit in this case is taken to be $\mathcal{R}_c M^2 \lesssim 0.1$, which corresponds to $\mathcal{R}_{c,1.4M_\odot} \lesssim 2.34 \times 10^{-8} $ m$^{-2}$}
		\footnotetext{Calculated from $\langle P \rangle M^2 - \Lambda$ relation.}
	\end{table*}

	Almost all curvature quantities generally show a consistent negative correlation with $I/M^3$ and $\Lambda$, except for the Ricci scalar. In both $\mathcal{R}_c M^2$ and $\langle \mathcal{R} \rangle M^2$, the relations exhibit local maxima at smaller values of $I/M^3$ and $\Lambda$, which correspond to larger masses. For both NS and QS, these values do not exceed $\sim10^{-1}$. This situation relates to the change in the $\varepsilon-3P$ term within $\mathcal{R}$. The dominance shifts from energy density to $3P$ in high-mass stars. These trace characteristics within $\mathcal{R}$ may be related to instabilities  \cite{Fujimoto2022,Ferreira2024}.
	
	Building on the previous discussion of curvature quantities, it is important to highlight that the $\mathcal{R}_c$ measurement obtained exclusively from GW170817 in Table \ref{Tab:DeducedQTYS} provides only a lower bound, as its distribution centers around the peak with large uncertainty. In such cases, treating $\mathcal{R}_{c,1.4M\odot} \lesssim 2.34 \times 10^{-8} \, \mathrm{m}^{-2}$ as an upper limit proves useful. This issue does not apply to the combined GW170817+LMXB measurements, which produce significantly tighter constraints.
	
	Continuing from the discussion of measurement constraints, the value of $\mathcal{K}_{surf}$ estimated in this study agrees with results from Ref. \cite{Das2023}. In their work, the "Surface Curvature (SC)" term is defined as $\text{SC}_{1.4} = {\sqrt{\mathcal{K}_{surf,1.4 M\odot}}}/{\sqrt{\mathcal{K}_{surf,\odot}}}$, where $\mathcal{K}_{surf,\odot}$ represents the surface Kretchmann curvature on the solar surface. For GW170817 data, we obtained $\text{SC}_{1.4} = 3.294^{+1.723}_{-1.310} \times 10^{14}$, while for GW170817+LMXB, we found $\text{SC}_{1.4} = 3.250^{+0.591}_{-0.509} \times 10^{14}$.
	
	Shifting focus from curvature to other canonical mass stellar properties, quantities such as central and average pressures, densities, and the central pressure-density ratio can also be derived from $\Lambda_{1.4}$ measurements. Our analysis shows that the GW170817 $\Lambda_{1.4}$ constraint yields $\log{10} (\varepsilon_{c,1.4}) = 15.16^{+0.21}_{-0.25}$ and $\log{10}(P_{c,1.4}) = 35.33^{+0.33}_{-0.37}$ in cgs units. These results partially overlap with multimessenger inference values from Ref. \cite{Rutherford2024}, which reported around $\log{10} (\varepsilon_{c,1.4}) \in [14.87,14.94]$ and $\log{10}(P_{c,1.4}) \in [34.96,35.07]$ across different models. The tighter GW170817+LMXB $\Lambda_{1.4}$ constraint produces $\log{10} (\varepsilon_{c,1.4}) = 15.16^{+0.08}_{-0.09}$ and $\log{10}(P_{c,1.4}) = 35.32^{+0.13}_{-0.13}$, slightly exceeding literature values. 
	
	Expanding on the discussion of central stellar properties, it is notable that, unlike central pressure cases, the $P_c/\varepsilon_c$ ratio is approximately consistent with the multimessenger inference result for both GW170817 and GW170817+LMXB. By calculating the canonical-mass central pressure to central energy density ratio, taking the modal values from each scenario set in \cite{Rutherford2024}, the equivalent values for this ratio in canonical-mass stars fall within $P_c/\varepsilon_c \in [0.137, 0.150]$, which overlap with the bounds deduced in this study, as shown in Tab. \ref{Tab:DeducedQTYS}.
	
	To synthesize these findings, we stress that the aforementioned bounds do not represent complete statistical constraints, as they are derived solely from $\Lambda_{1.4}$ measurements. A comprehensive error analysis would require a broader EoS parameter space and quantification of systematic uncertainties, potentially resulting in wider but statistically consistent uncertainty ranges.

	\section{Conclusion}
	\label{SECT:conclusion}
	
	An incomplete understanding of the equation of state at extremely high densities introduces additional uncertainty when linking observations to the actual properties of stars. In this study, we connect measurable quantities—tidal deformability and moment of inertia—to the structure of compact, highly relativistic stars. We examine how scalar curvatures in general relativity scale and relate these curvatures to the known connection between a star's dimensionless moment of inertia and tidal deformability in neutron stars and quark stars. Our results provide a framework for constraining the properties of ultra-dense matter and enhancing the interpretation of astrophysical observations. Our analysis reveals that these curvature-based scaling relations strengthen the predictive power of universal relations for both neutron stars and quark stars, providing more robust connections between observables and compact star interiors. 
	
	The well-known $I-\Lambda$ relation is a cornerstone of universality. Building on this, we found that scalar curvature quantities such as Ricci scalar ($\mathcal{R}$), contraction of the Ricci tensor ($\mathcal{J}$), Kretschmann scalar ($\mathcal{K}$), and Weyl scalar ($\mathcal{W}$) show noticeable correlation with tidal deformability and moment of inertia. Importantly, our results reveal that universality affects not only the global values of these quantities (such as volume-averaged scalars) but also local values, including central and surface measurements. However, we observe that volume-averaged quantities exhibit a stronger correlation than their central or surface counterparts. Furthermore, we found that the Ricci scalar, represented by the normalized central $\mathcal{R}_c M^2$ and average $\langle\mathcal{R}\rangle M^2$ scalars, has an approximately EoS-independent maximum value around $\lesssim 0.1$. We expect this limit to be tied to the trace anomaly characteristics of a compact star. Finally, using the above scaling relations, we show that the scalar curvature values of a canonical mass of NS can be estimated from $\Lambda_{1.4}$ measurements.
	
	In this study, we also demonstrate that the stellar interior variables, which constitute scalar curvature quantities, exhibit a strong universality with $I$ and $\Lambda$. Normalizing these quantities by mass with the same treatment as scalar curvatures, as well as their combinations, we found several new universal relations; some of them have already been discovered in the literature. Our main findings on this aspect are:
	\begin{enumerate}
		\item The central pressure $P_c M^2$ and central energy density $\varepsilon_c M^2$ exhibit a strong correlation with $I$ and $\Lambda$, with overlapping trends for NS and QS, similar to their ratio of $P_c/\varepsilon_c$ in which is already well-known in the literature. These universal relations enable central pressure and central energy density estimation of $1.4 M_\odot$ stars directly from $\Lambda_{1.4}$ measurements, which we found $\varepsilon_{c,1.4} = 802.1846^{+171.4175}_{-142.7433}$ MeV fm$^{-3}$ and $P_{c,1.4} = 257.7262^{+46.8461}_{-40.3749}$ MeV fm$^{-3}$ from GW170817+Low Mass X-ray Binary constraints. These values are consistent with typical values deduced using other methods, such as Bayesian inference, throughout a set of EoSs.  
		\item We establish a strong, universal relation between certain observational quantities and volume-averaged interior properties of neutron stars. Using a similar approach to the previously mentioned case, the volume-averaged values for canonical mass neutron stars can also be determined to be $\langle \varepsilon \rangle_{1.4} = 257.7262^{+46.8461}_{-40.3749}$ MeV fm$^{-3}$ and $\langle P \rangle_{1.4} = 20.8001^{+6.4991}_{-4.9553}$ MeV fm$^{-3}$. Meanwhile, the ratio between average pressure and density ($\langle P \rangle/\langle \varepsilon \rangle$) exhibits very strong universality with overlapping neutron star and quark star trends. 
	\end{enumerate}  
	
	The results in this study require further work to achieve statistical robustness. To achieve this, a broader set of EoSs covers varied degrees of freedom is required. Also, measure trend dispersion, for example, by estimating heteroskedastic errors based on the size of each quantity. Taking these steps helps account for uncertainties arising from measurement errors, EoS variation, and fitting accuracy in the error estimation. This process will implies a more statistically meaningful distribution of parameters values.
	
	All calculations are performed using general relativity, making the results valid only within this framework. Extending the analysis to alternative and modified gravity theories would reveal how geometric quasi-universal relations change or persist, which may indicate how strong-field gravitational behavior depends on the underlying theory. This could also reduce reliance on EoS uncertainties by examining the effect of different curvature definitions on universality and identifying which observables could most effectively constrain these relations.

	
	\begin{acknowledgments}
		The author thanks P. T. H. Pang for pointing out the surface condition of quark stars in the tidal deformability calculations.
	\end{acknowledgments}
	
	\appendix
	\section{Fitted polynomial coefficients}
	We provide fitting coefficient for quantities discussed in this study in Table 	\ref{tab:fitting_coeff_NS_Lam}--\ref{tab:fitting_coeff_QS_I}.
	
	\begin{table}
		\caption{Polynomial fitting coefficients $a_0$–$a_4$ for the relations between stellar quantities $Y$ and tidal deformability $\Lambda$ of neutron star equation of states, along with the relative residuals in logarithmic and linear scales, $\mu_{(\ln Y)}$ and $\mu_{(Y)}$.}
		\begin{tabular}{ccccccrr}
			\hline\hline
			\multicolumn{8}{c}{NS, $X = \Lambda$} \\ \hline
			Y & $a_0$ & $a_1$ & $a_2$ & $a_3$ & $a_4$ & $\mu_{(\ln Y)}$ (\%) & $\mu_{(Y)}$ (\%) \\
			\hline
			$\mathcal{R}_c M^2$ & -1.215e+01 & 4.996e+00 & -9.014e-01 & 6.391e-02 & -1.650e-03 & 5.683936 & 19.689331 \\
			$\mathcal{K}_c M^4$ & 1.102e+00 & -5.867e-01 & -9.040e-02 & 7.050e-03 & -1.982e-04 & 33.873261 & 26.842332 \\
			$\mathcal{J}_c M^4$ & 5.389e-01 & -6.851e-01 & -6.593e-02 & 5.001e-03 & -1.401e-04 & 16.345294 & 26.856776 \\
			$\langle \mathcal{R} \rangle M^2$ & -4.675e+00 & 1.237e+00 & -2.896e-01 & 2.161e-02 & -6.286e-04 & 2.793495 & 10.129984 \\
			$\langle \mathcal{K} \rangle M^4$ & -9.918e-01 & -6.812e-01 & -2.819e-02 & 1.089e-03 & -5.597e-05 & 0.669477 & 3.198004 \\
			$\langle \mathcal{J} \rangle M^4$ & -1.690e+00 & -8.166e-01 & 6.858e-03 & -2.221e-03 & 4.302e-05 & 0.840192 & 5.119249 \\
			$\langle \mathcal{W} \rangle M^4$ & -3.563e+00 & -1.313e-01 & -1.416e-01 & 1.131e-02 & -3.661e-04 & 1.967358 & 10.744741 \\
			$\mathcal{W}_{\text{surf}} M^4$ & \multirow{2}*{-2.244e+00} & \multirow{2}*{-7.506e-01} & \multirow{2}*{1.043e-02} & \multirow{2}*{-3.098e-03} & \multirow{2}*{4.473e-05} & \multirow{2}*{2.578728} & \multirow{2}*{14.928471 } \\
			$\mathcal{K}_{\text{surf}} M^4$ &  &  &  &  &  &  \\ \hline
			$\epsilon_c M^2$ & -3.561e+00 & -7.987e-02 & -7.651e-02 & 5.664e-03 & -1.539e-04 & 2.634863 & 13.429373 \\
			$P_c M^2$ & -3.590e+00 & -5.386e-01 & -4.554e-02 & 3.951e-03 & -1.188e-04 & 1.655931 & 10.495507 \\
			$P_c/\epsilon_c$ & -2.976e-02 & -4.588e-01 & 3.097e-02 & -1.713e-03 & 3.507e-05 & 2.709562 & 3.565917 \\
			$\langle \epsilon \rangle M^2$ & -4.490e+00 & -3.753e-01 & 5.213e-03 & -1.549e-03 & 2.236e-05 & 1.144881 & 7.450781 \\
			$\langle P \rangle M^2$ & -5.308e+00 & -7.650e-01 & 2.071e-02 & -2.076e-03 & 2.714e-05 & 0.701229 & 6.337811 \\
			$\langle P \rangle R^3/M$ & -2.250e+00 & -3.896e-01 & 1.550e-02 & -5.268e-04 & 4.780e-06 & 0.358289 & 1.216032 \\
			\hline\hline
		\end{tabular}
		\label{tab:fitting_coeff_NS_Lam}
	\end{table}
	
	\begin{table}
		\caption{Polynomial fitting coefficients $a_0$–$a_4$ for the relations between stellar quantities $Y$ and tidal deformability $\Lambda$ of quark star equation of states, along with the relative residuals in logarithmic and linear scales, $\mu_{(\ln Y)}$ and $\mu_{(Y)}$.}
		\begin{tabular}{ccccccrr}
			\hline\hline
			\multicolumn{8}{c}{QS, $X = \Lambda$} \\ \hline
			Y & $a_0$ & $a_1$ & $a_2$ & $a_3$ & $a_4$ & $\mu_{(\ln Y)}$ (\%) & $\mu_{(Y)}$ (\%) \\
			\hline
			$\mathcal{R}_c M^2$ & -7.002e+00 & 2.273e+00 & -4.093e-01 & 2.582e-02 & -5.893e-04 & 5.336851 & 19.134144 \\
			$\mathcal{K}_c M^4$ & 5.442e+00 & -3.050e+00 & 3.271e-01 & -2.202e-02 & 5.125e-04 & 79.010602 & 15.446416 \\
			$\mathcal{J}_c M^4$ & 4.762e+00 & -3.069e+00 & 3.337e-01 & -2.250e-02 & 5.238e-04 & 30.825140 & 15.537161 \\
			$\langle \mathcal{R} \rangle M^2$ & -3.791e+00 & 7.762e-01 & -1.670e-01 & 9.651e-03 & -2.097e-04 & 1.305328 & 4.238892 \\
			$\langle \mathcal{K} \rangle M^4$ & -3.553e-01 & -1.014e+00 & 4.178e-02 & -4.435e-03 & 1.193e-04 & 0.997105 & 3.605974 \\
			$\langle \mathcal{J} \rangle M^4$ & -1.570e+00 & -7.744e-01 & 9.111e-03 & -2.459e-03 & 7.570e-05 & 0.585534 & 2.740996 \\
			$\langle W^2 \rangle M^4$ & -1.190e-01 & -2.183e+00 & 1.279e-01 & -8.994e-03 & 2.117e-04 & 2.653670 & 21.198748 \\
			$\mathcal{W}_{\text{surf}} M^4$ & \multirow{2}*{-2.988e+00} & \multirow{2}*{-1.962e-02} & \multirow{2}*{-1.086e-01} & \multirow{2}*{5.201e-03} & \multirow{2}*{-1.009e-04} & \multirow{2}*{0.804274} & \multirow{2}*{3.477643} \\
			$\mathcal{K}_{\text{surf}} M^4$ &  &  &  &  &  &  &  \\
			\hline
			$\epsilon_c M^2$ & -1.139e+00 & -1.442e+00 & 1.556e-01 & -1.064e-02 & 2.494e-04 & 1.600239 & 7.504139 \\
			$P_c M^2$ & -1.771e+00 & -1.660e+00 & 1.558e-01 & -1.062e-02 & 2.492e-04 & 1.163872 & 7.191231 \\
			$P_c/\epsilon_c$ & -6.324e-01 & -2.184e-01 & 2.233e-04 & 1.549e-05 & -1.676e-07 & 0.818119 & 1.202532 \\
			$\langle \epsilon \rangle M^2$ & -4.862e+00 & -9.809e-03 & -5.432e-02 & 2.601e-03 & -5.045e-05 & 0.305966 & 1.740877 \\
			$\langle P \rangle M^2$ & -5.531e+00 & -5.166e-01 & -1.572e-02 & 3.342e-04 & -8.577e-07 & 0.174726 & 1.372451 \\
			$\langle P \rangle R^3/M$ & -2.101e+00 & -5.068e-01 & 3.860e-02 & -2.266e-03 & 4.959e-05 & 0.308173 & 1.115660 \\
			\hline\hline
		\end{tabular}
	\end{table}
	
	\begin{table}
		\caption{Polynomial fitting coefficients $a_0$–$a_4$ for the relations between stellar quantities $Y$ and dimensionless moment of inertia $I/M^3$ of neutron star equation of states, along with the relative residuals in logarithmic and linear scales, $\mu_{(\ln Y)}$ and $\mu_{(Y)}$.}
		\begin{tabular}{ccccccrr}
			\hline\hline
			\multicolumn{8}{c}{NS, $X = I/M^3$} \\ \hline
			Y & $a_0$ & $a_1$ & $a_2$ & $a_3$ & $a_4$ & $\mu_{(\ln Y)}$ (\%) & $\mu_{(Y)}$ (\%) \\
			\hline
			$\mathcal{R}_c M^2$ & -3.937e+01 & 4.458e+01 & -1.928e+01 & 3.484e+00 & -2.300e-01 & 6.321996 & 21.868873 \\
			$\mathcal{K}_c M^4$ & 1.811e+01 & -1.768e+01 & 5.343e+00 & -8.885e-01 & 5.601e-02 & 54.914619 & 24.918617 \\
			$\mathcal{J}_c M^4$ & 1.836e+01 & -1.900e+01 & 6.001e+00 & -1.020e+00 & 6.527e-02 & 15.306483 & 24.956242 \\
			$\langle \mathcal{R} \rangle M^2$ & -1.671e+01 & 1.893e+01 & -9.019e+00 & 1.709e+00 & -1.195e-01 & 3.079220 & 11.174960 \\
			$\langle \mathcal{K} \rangle M^4$ & 1.263e+01 & -1.438e+01 & 4.366e+00 & -7.453e-01 & 4.565e-02 & 1.099730 & 4.848858 \\
			$\langle \mathcal{J} \rangle M^4$ & 1.264e+01 & -1.566e+01 & 5.037e+00 & -8.896e-01 & 5.631e-02 & 1.469013 & 7.789223 \\
			$\langle \mathcal{W} \rangle M^4$ & 5.536e+00 & -7.682e+00 & 1.348e+00 & -1.470e-01 & 2.921e-03 & 1.765947 & 9.516652 \\
			$\mathcal{W}_{\text{surf}} M^4$ & \multirow{2}*{8.794e+00} & \multirow{2}*{-1.159e+01} & \multirow{2}*{3.371e+00} & \multirow{2}*{-5.876e-01} & \multirow{2}*{3.424e-02} & \multirow{2}*{3.132072} & \multirow{2}*{17.158042 } \\
			$\mathcal{K}_{\text{surf}} M^4$&  &  &  &  &  &  \\ \hline
			$\epsilon_c M^2$ & 1.839e+00 & -4.700e+00 & 9.493e-01 & -1.306e-01 & 7.036e-03 & 2.525134 & 12.832190 \\
			$P_c M^2$ & 1.020e+01 & -1.488e+01 & 4.772e+00 & -8.044e-01 & 5.104e-02 & 1.438147 & 9.163696 \\
			$P_c/\epsilon_c$ & 8.357e+00 & -1.018e+01 & 3.823e+00 & -6.738e-01 & 4.400e-02 & 3.407274 & 4.351982 \\
			$\langle \epsilon \rangle M^2$ & 1.029e+00 & -5.793e+00 & 1.685e+00 & -2.938e-01 & 1.712e-02 & 1.352803 & 8.581032 \\
			$\langle P \rangle M^2$ & 7.920e+00 & -1.492e+01 & 5.010e+00 & -8.716e-01 & 5.457e-02 & 1.020864 & 8.795722 \\
			$\langle P \rangle R^3/M$ & 5.458e+00 & -9.126e+00 & 3.324e+00 & -5.778e-01 & 3.745e-02 & 0.477409 & 1.782728 \\
			\hline\hline
		\end{tabular}
	\end{table}
	
	\begin{table}
		\caption{Polynomial fitting coefficients $a_0$–$a_4$ for the relations between stellar quantities $Y$ and dimensionless moment of inertia $I/M^3$ of quark star equation of states, along with the relative residuals in logarithmic and linear scales, $\mu_{(\ln Y)}$ and $\mu_{(Y)}$.}
		\begin{tabular}{ccccccrr}
			\hline\hline
			\multicolumn{8}{c}{QS, $X = I/M^3$} \\ \hline
			Y & $a_0$ & $a_1$ & $a_2$ & $a_3$ & $a_4$ & $\mu_{(\ln Y)}$ (\%) & $\mu_{(Y)}$ (\%) \\
			\hline
			$\mathcal{R}_c M^2$ & -1.854e+01 & 1.843e+01 & -7.491e+00 & 1.199e+00 & -6.920e-02 & 6.182700 & 21.392305 \\
			$\mathcal{K}_c M^4$ & 3.219e+01 & -3.243e+01 & 1.053e+01 & -1.638e+00 & 9.287e-02 & 80.788845 & 21.793134 \\
			$\mathcal{J}_c M^4$ & 3.149e+01 & -3.251e+01 & 1.060e+01 & -1.653e+00 & 9.381e-02 & 15.887738 & 21.991321 \\
			$\langle \mathcal{R} \rangle M^2$ & -8.074e+00 & 7.427e+00 & -3.342e+00 & 5.324e-01 & -3.062e-02 & 2.050320 & 6.708764 \\
			$\langle \mathcal{K} \rangle M^4$ & 1.025e+01 & -1.103e+01 & 2.799e+00 & -4.299e-01 & 2.421e-02 & 1.260449 & 4.871265 \\
			$\langle \mathcal{J} \rangle M^4$ & 7.058e+00 & -8.436e+00 & 1.869e+00 & -2.851e-01 & 1.601e-02 & 0.706246 & 3.351905 \\
			$\langle \mathcal{W} \rangle M^4$ & 2.316e+01 & -2.558e+01 & 7.251e+00 & -1.084e+00 & 5.989e-02 & 2.674967 & 21.972731 \\
			$\mathcal{W}_{\text{surf}} M^4$ & \multirow{2}*{3.859e-01} & \multirow{2}*{-1.135e+00} & \multirow{2}*{-8.765e-01} & \multirow{2}*{1.544e-01} & \multirow{2}*{-9.387e-03} & \multirow{2}*{0.930811} & \multirow{2}*{4.074717 } \\
			$\mathcal{K}_{\text{surf}} M^4$ &  &  &  &  &  &  \\ \hline
			$\epsilon_c M^2$ & 1.138e+01 & -1.518e+01 & 4.928e+00 & -7.697e-01 & 4.376e-02 & 2.043879 & 10.248396 \\
			$P_c M^2$ & 1.355e+01 & -1.801e+01 & 5.612e+00 & -8.634e-01 & 4.865e-02 & 1.564536 & 10.389318 \\
			$P_c/\epsilon_c$ & 2.169e+00 & -2.827e+00 & 6.842e-01 & -9.377e-02 & 4.884e-03 & 0.927322 & 1.349359 \\
			$\langle \epsilon \rangle M^2$ & -3.175e+00 & -5.675e-01 & -4.383e-01 & 7.722e-02 & -4.694e-03 & 0.355284 & 2.040191 \\
			$\langle P \rangle M^2$ & 1.394e+00 & -6.567e+00 & 1.377e+00 & -1.914e-01 & 1.005e-02 & 0.222627 & 1.771989 \\
			$\langle P \rangle R^3/M$ & 3.136e+00 & -6.000e+00 & 1.815e+00 & -2.687e-01 & 1.475e-02 & 0.574889 & 2.212071 \\
			\hline\hline
		\end{tabular}
		\label{tab:fitting_coeff_QS_I}
	\end{table}

	\bibliography{apssamp} 
	
\end{document}